\documentclass[12pt]{article}
\usepackage[utf8]{inputenc}
\usepackage{amsmath, amssymb}
\usepackage{geometry}
\usepackage{multirow}
\usepackage{tikz}
\usepackage{dsfont}
\usepackage[normalem]{ulem}
\usepackage{verbatim}
\usetikzlibrary{arrows.meta, positioning}
\usepackage{ytableau}
 \newcommand{\bea}{\begin{eqnarray}}
\newcommand{\eea}{\end{eqnarray}}
\usepackage{listings}     
\usepackage{xcolor}    
\usepackage[numbers,sort&compress]{natbib}
\usepackage{url}

\begin{document}
\newcommand\mytitle{The Hilbert Series and the Flavor Invariants of the 3HDM   }
\newcommand\mypreprint{UCI-TR-2026-03}
\begin{titlepage}
	\begin{flushright}
		\mypreprint
	\end{flushright}
	
	\vspace*{2cm}
	
	\begin{center}
		{\Large\sffamily\bfseries\mytitle}
		
		\vspace{1cm}
		
		\renewcommand*{\thefootnote}{\fnsymbol{footnote}}
		
		\textbf{%
			Eric Bryan\footnote{ebryan2@uci.edu}, 
			Arvind Rajaraman\footnote{arajaram@uci.edu}
		}
		\\[8mm]
		\textit{\small
			~Department of Physics and Astronomy\\ University of California\\ Irvine, CA 92697-4575, USA
		}
	
	\end{center}

	\vspace*{1cm}
	\begin{abstract}
We perform a systematic study of invariant operators in the three-Higgs-doublet model (3HDM). We compute the closed-form multigraded Hilbert series associated with the global symmetry group of the theory. In addition, we construct explicit expressions for the corresponding invariant operators up to cubic order in the couplings. As a phenomenological application, we further organize the invariants up to quadratic order in the field content into CP-even and CP-odd combinations. For the special case of invariants involving no $\mathbf{27}$ representations, our results are complete to all orders. 
\end{abstract}
	\vspace*{1cm}
\end{titlepage}

\setcounter{footnote}{0}

\section{Introduction}\label{3hdmlag}
The $N$-Higgs-doublet-models (NHDM) are natural extensions of the Standard Model that introduce  additional Higgs doublets~\cite{Nishi:2006tg,Nishi:2007nh,Ferreira:2008zy,Ivanov:2010ww,Grzadkowski:2009bt}.
These are interesting because they can supply  new sources for CP violation, possible dark matter candidates, and  novel new patterns of electroweak symmetry breaking.  

The simplest realization of this framework is the two-Higgs-doublet model (2HDM),
%.
%The model has also 
which has been studied extensively as a minimal extension of the Standard Model scalar sector. In the 2HDM, the enlarged Higgs potential allows for both explicit and spontaneous CP violation~\cite{Branco:2011iw,Ginzburg:2004vp,Gunion:2005ja,Botella:1994cs,Lavoura:1994fv,Maniatis:2007vn,Haber:2006ue,Khater:2003ym,Barroso:2007rr}. The 2HDM has also been explored as a framework for generating neutrino masses, either through extended Yukawa structures or via radiative mechanisms that naturally suppress the neutrino mass scale~\cite{Antusch:2001vn,Atwood:2005bf,Liu:2016mpf,Cheung:2017lpv,Arcadi:2017wqi,Camargo:2019ukv}. Depending on the imposed symmetries, the model can also accommodate stable scalar particles that serve as viable dark matter candidates \cite{Camargo:2019ukv,LopezHonorez:2006gr,Gustafsson:2007pc,Dolle:2009fn,Chao:2012pt,Goudelis:2013uca,LopezHonorez:2010eeh,LopezHonorez:2010tb,Arhrib:2013ela,Bonilla:2014xba,Queiroz:2015utg,Arcadi:2018pfo}. Moreover, extended Higgs sectors exhibit nontrivial symmetry-breaking patterns, including electroweak and soft symmetry breaking, that lead to interesting low-energy symmetry
structures~\cite{Maniatis:2006fs,Deshpande:1977rw,Ferreira:2022gjh}.
{The Minimal Supersymmetric Standard Model (MSSM) also requires a 2HDM stucture~\cite{Haber:1984rc}, where 
a second Higgs doublet is required to ensure anomaly cancellation and to provide masses separately to up- and down-type fermions.}

{However, it was found that the minimal 2HDM cannot simultaneously accommodate explicit CP violation and a viable dark matter candidate~\cite{Keus:2016orl}. This motivates consideration of the next simplest multi-Higgs-doublet extension.}
In this work, we consider the three-Higgs-doublet model (3HDM), which constitutes the next nontrivial extension beyond the extensively studied two-Higgs-doublet model. The introduction of a third scalar doublet substantially enlarges the scalar potential and leads to a richer symmetry structure. Consequently, the 3HDM admits a wide array of discrete and continuous global symmetries, together with intricate patterns of spontaneous symmetry breaking that are inaccessible in simpler Higgs sectors. These features give rise to new sources of CP violation, viable dark matter candidates, and novel possibilities for organizing flavor symmetries of quarks and leptons.

{Studies of the 3HDM~\cite{Ivanov:2012fp,Ivanov:2012ry,Kuncinas:2025uty,Keus:2013hya,deMedeirosVarzielas:2019rrp}
have classified the symmetry groups realizable in the scalar potential of the 3HDM. Such symmetries play an important role in determining the phenomenology
of multi-Higgs models, including the structure of the scalar spectrum, CP
properties, and the pattern of symmetry breaking.} It has also been shown that symmetries acting on the three Higgs doublets may provide a natural framework for describing the observed flavor structure of the Standard Model, and may offer a possible connection to the existence of three generations of quarks and leptons~\cite{Keus:2013hya,GonzalezFelipe:2013xok}. 

{Previous authors~\cite{Ivanov:2014doa,Emmanuel-Costa:2017zkm,Ivanov:2018ime,deMedeirosVarzielas:2017vph} have investigated CP violation in a variety of 3HDM scenarios, including models
with imposed discrete symmetries. By analyzing the symmetries of the scalar
potential and the vacuum structures that arise in these models, they identified
regions of parameter space exhibiting either explicit or spontaneous CP
violation.} There has also been work extending symmetry-based analyses of the 3HDM to investigate viable scalar dark matter candidates~\cite{Kuncinas:2022hwv,Kuncinas:2024zjq,Kuncinas:2025mcn,Deng:2025dcq,Sakurai:2022roq,Farhan:2025vui,DuttaBanik:2025fce}. These studies demonstrate how residual symmetries—such as $S_3$ or unbroken $U(1)$ symmetries—can stabilize dark matter within the 3HDM.

Many of the above methods rely on classifying symmetries of the 3HDM and the parameters which are invariant under such symmetries. Therefore a central challenge in the study of extended Higgs sectors is the identification of physical parameters that are independent of basis choices and field redefinitions. Indeed, all observable quantities—such as CP-violating observables analogous to the Jarlskog invariant, as well as invariant combinations appearing in renormalization group equations~\cite{Ivanov:2005hg,Ivanov:2006yq,Bednyakov:2018cmx}—must ultimately be expressible in terms of operators that are invariant under the full set of local and global symmetries of the theory. 

In the case of the 2HDM, this problem has been addressed systematically. A minimal generating set of invariant operators was identified in~\cite{Bento:2021hyo,Bento:2020jei,Trautner:2018ipq,Davidson:2005cw,Ivanov:2019kyh,Bednyakov:2025sri}. These approaches made use of bilinears and birdtrack techniques to construct the invariants explicitly. Such methods enabled researchers to address important questions about the 2HDM, for example the identification of CP-odd combinations of physical observables.
However, a similar systematic study has not yet been carried out for the 3HDM, primarily due to the much larger parameter space and the increased complexity of the 
global symmetry and its irreducible representation content. 

In this work, we aim to perform a similar analysis for the 3HDM. {We present three main results. First, we compute the complete closed-form multigraded Hilbert series of the 3HDM, which encodes both the number of independent invariant operators and their representation content.} As will be described in the following sections, the Hilbert series provides a compact and powerful way to organize invariant operators. Closely related techniques appear in a variety of other physical contexts~\cite{Jenkins:2009dy}. For example, in $\mathcal{N}=1$ supersymmetric QCD, Hilbert series have been used to characterize and match the chiral rings of theories related by Seiberg duality~\cite{Pouliot:1998yv,Seiberg:1994pq,Hanany:2008sb}. Furthermore, these techniques have also been employed in model building to systematically construct Lagrangians consistent with imposed symmetries~\cite{Lehman:2015via}.

However, Hilbert series are in general difficult to compute for large representations composed of many irreducible components, such as those appearing in the 3HDM. 
 As will be shown below, the task of computing the full Hilbert series of the 3HDM is technically challenging due to the appearance of higher-order poles, branch-cut ambiguities arising in the contour integrals, and computational difficulties associated with summing large polynomials with large coefficients.
Indeed, previous studies of the 3HDM Hilbert series were restricted either to power-series expansions of the integrand or to the computation of the ungraded Hilbert series~\cite{Bento:2021hyo}, {which made use of $\Omega$ calculus and Machon's partition analysis~\cite{Xin:2004}. The ungraded Hilbert series is obtained by identifying all grading variables of the multigraded Hilbert series, thereby counting invariant operators only according to their total degree. While this provides the total number of invariants at a given order, it does not retain information about the representation content from which those invariants are constructed.}  
 One of the main results of this work is the computation of the closed form multigraded Hilbert series of the 3HDM.

In addition to counting invariant operators using the Hilbert series, {our second result resides in  constructing the basis of invariant operators} explicitly. To this end, we employ a method based on background field techniques. These methods were originally developed and proven in the context of supersymmetric theories~\cite{Luty:1995sd,Brax:2001an}, but have since found broader application. The basic idea is to 
treat some parameters (such as the masses) as background values that partially break the global symmetry group.  The set of invariants under this reduced symmetry can be classified more easily. One then combines these invariants with the background fields to reconstruct the invariants of the full symmetry group. As we show, this can be  done efficiently by generating  candidate invariants of the full theory and successively matching them to the invariants in the theory with reduced symmetry.  This approach allows one to systematically build a basis of invariant operators. Variants of this method have been successfully applied in a number of related contexts~\cite{Berger:2018dxg,Almumin:2020yoq,Rajaraman:2022zah}.

As will be discussed in the following section, the 3HDM contains three adjoint representations and one $\mathbf{27}$ representation of a global flavor symmetry group $SU(3)$. While in principle one can compute all independent invariants of the theory, their sheer number makes it impractical to present the full set explicitly. Instead, we restrict our attention to operators up to cubic order in the quartic couplings. For the special case of invariants containing no $\mathbf{27}$ fields, however, we determine the generating set to arbitrary order in the quartic couplings. We expect this truncation to be sufficient for phenomenological applications.

{Our third and final result is the application of these invariants in a phenomenological context. In particular, we use our basis of invariants up to cubic order in the couplings to begin the construction of CP-even and CP-odd basis invariants. We explicitly identify CP-even and CP-odd combinations for invariants involving up to four fields and discuss how the construction may be systematically extended to higher orders.
}

Our paper is organized as follows. 
In Section~\ref{3hdmoverview}, we provide an overview of the 3HDM Lagrangian and identify the building blocks required to construct invariant operators. In Section~\ref{hseries}, we introduce the Hilbert series and  
present the explicit formula used to compute the Hilbert series for the 3HDM. In Section~\ref{hseriesCalc}, we describe the steps involved in performing this calculation, emphasizing the technical obstacles that previously prevented its completion and explaining how each of these challenges can be overcome. In Section~\ref{HSResults}, we present the results of the Hilbert series calculation.

We then turn to the explicit construction of the independent invariant operators of the 3HDM. In Section~\ref{Invariants}, we describe our approach and present the calculation of the lowest-order invariants of the theory.  Section~\ref{invariantresults} presents the resulting basis of invariant operators through cubic order in the couplings. {In Section~\ref{CPinvs}, we apply these results in a phenomenological context by constructing CP-even and CP-odd invariants up to quartic order in the field content.}
Finally, we close with a discussion of our results.

\section{Overview of the 3HDM}\label{3hdmoverview}
  The most general renormalizable scalar potential of the  NHDM contains quadratic and quartic terms involving the $N$ new Higgs fields. It may be written as
\begin{align}
    V_H = (\mu^2)_{i}^{~j}(\Phi^\dag)^i \Phi_j + \lambda_{ij}^{~~kl}\big((\Phi^\dag)^i \Phi_k\big)\big((\Phi^\dag)^j \Phi_l\big)
\end{align}
where the Higgs fields are denoted by $\Phi_i$, with $i=1,2,\dots,N$. Throughout, the $SU(2)_L$ gauge indices are suppressed {and we assume $SU(2)_L$ scalar doublets.} {The introduction of these new terms to the Standard Model Lagrangian induces a global $SU(N)$ symmetry under which the Higgs fields $\Phi_i$ transform as fundamentals. We have adopted the convention that fundamental indices are written with lower indices (subscripts), while anti-fundamental indices are written with upper indices (superscripts).} Since the center of $SU(N)$ acts trivially on the gauge-invariant bilinears $(\Phi^\dag)^i\Phi_j$, the faithful symmetry group is $PSU(N)=SU(N)/Z_N$. Nevertheless, we will continue to use the standard $SU(N)$ representation labels when discussing the representation content of the theory.  {One may also note that although the Higgs fields transform in the fundamental representation of $SU(N)$, the scalar potential depends only on the gauge-invariant bilinears, which transform in the representation $\bar{\bf{N}}\otimes \bf{N} = \bf{1}\oplus \bf{Adj}$. Consequently, questions concerning basis transformations, symmetries, and basis-invariant observables are naturally formulated in terms of the singlet and adjoint representations rather than the Higgs fields themselves.}

{The NHDM contains ${(N^4+N^2+2)/}{2}$ physical parameters~\cite{Olaussen:2010aq}}. Subsequently the 3HDM contains a large number of free parameters and due to the large gauge symmetry, this makes the identification of physical observables particularly nontrivial. All observables must ultimately be expressible in terms of the  bilinear parameters $(\mu^2)_{i}^{~j}$ and quartic couplings $\lambda_{ij}^{~~kl}$, subject to invariance under the global ${SU}(3)$ flavor symmetry. It is therefore natural to begin by classifying these parameters according to their transformation properties under ${SU}(3)$.

Due to the Hermiticity of the scalar potential and {the symmetry of the quartic interaction under interchange of the two gauge-invariant bilinears}, the couplings $\mu$ and $\lambda$ satisfy the relations
\begin{align}
(\mu^2)_{i}^{~j}=((\mu^2)_j^{~i})^{*}, \qquad 
\lambda_{ij}^{~~kl}=\lambda_{ji}^{~~lk}=(\lambda_{kl}^{~~ij})^{*}\,.
\end{align}
Under the ${SU}(3)$ action, the mass parameters $(\mu^2)_{i}^{~j}$ transform in the representation
\begin{align}
    \bar{\mathbf{3}} \otimes \mathbf{3}
    &= \mathbf{1} \oplus \mathbf{8}\, .
\end{align}
The singlet corresponds to the trace component,
$(\mu^2)_i^{~i}$,
while the traceless part transforms in the adjoint representation.

{Since the quartic potential is written in terms of gauge-invariant bilinears
$(\Phi^\dag)^i\Phi_j$,
the $SU(2)_L$ gauge indices have already been contracted. The remaining representation-theoretic analysis therefore concerns only the flavor indices transforming under $SU(3)$.}
In order for $\lambda$ to satisfy the required symmetry properties, its irreducible components must have their fundamental and anti-fundamental indices either both symmetrized or both antisymmetrized. If they are simultaneously symmetrized, the representation is
\begin{align}
\lambda_{(ij)}^{~~(kl)}\rightarrow
%\big(
\mathbf{6}\otimes\bar{\mathbf{6}}
%\big)
=
%\big(
\mathbf{1}\oplus\mathbf{8}\oplus\mathbf{27}\,
%\big)
.
\end{align}
If they are simultaneously antisymmetrized, the representation is
\begin{align}
\lambda_{[ij]}^{~~[kl]}\rightarrow
%\big(
\mathbf{3}\otimes\bar{\mathbf{3}}
%\big)
=
%\big(
\mathbf{1}\oplus\mathbf{8}\,
%\big)
.
\end{align}

{Combining the representation content of the bilinear parameters $\mu^2$ and the quartic couplings $\lambda$, the scalar potential is parameterized by}
\begin{align}
\mathbf{1}\oplus\mathbf{1}\oplus\mathbf{1}\oplus\mathbf{8}\oplus\mathbf{8}\oplus\mathbf{8}\oplus\mathbf{27}\,.
\end{align}
We are therefore interested in constructing invariant structures built from the $\mathbf{8}$ and $\mathbf{27}$ representations of {${SU}(3)$}. {We will explicitly construct these representations in Section~\ref{Invariants}  but this is not needed for the calculation of the Hilbert series.
Therefore,} we will begin first  by constructing the Hilbert series for this representation content.

\section{Hilbert Series}\label{hseries}
In this section, we describe the basic formulae for  the Hilbert series.
Our discussion here will be brief and oriented toward
physical applications; one can see~\cite{Bento:2021hyo,Jenkins:2009dy,Trautner:2018ipq,Pouliot:1998yv,Hanany:2008sb} for more detailed treatments of the Hilbert series.

The Hilbert series is  defined as the power series
\begin{align}
H(t) = \sum_{k=0}^{\infty} d_k\, t^k\,,
\end{align}
where $d_k$ is the number of independent invariant operators at order $k$ {and $t$ is a fugacity that grades the operators according to their degree.}

Although the Hilbert series is defined as an infinite power series, it can in fact
be expressed 
as the power series expansion about $t=0$ of a rational function
$N(t)/D(t)$. 
In this convention, the Hilbert series is not unique: multiplying both the
numerator and denominator by an arbitrary polynomial in $t$ leaves the power
series expansion unchanged.

To compute the Hilbert series explicitly, one may use a set of standard results
from invariant theory. We state these formulae here without proof. For a finite group $G$, the Hilbert series is given by the {Molien formula}
\begin{align}
    H(t)={}&\frac{1}{|G|}\sum_{g\in G}\frac{1}{det(1-t\rho(g))}
\end{align}
where $\rho(g)$ denotes the representation of the group element $g$ acting on
the fields. This expression admits a natural generalization to continuous groups,
where the discrete sum is replaced by an integral over the group,
\begin{align}
    H(t)={}&\int_G d \mu_G\frac{1}{det(1-t\rho(g))}
\end{align}
where $d\mu_G$ is the Haar measure on $G$. {To compute the integrand one makes use of the Plethystic exponential which is defined as 
\begin{align}
    \mathrm{PE}[z_j,t,\bf{r}]&= \mathrm{exp}\left(\sum_{k\geq 1}\frac{t^k \chi_{\bf{r}}(z_j^k)}{k}\right)
\end{align}
where $\chi_{\bf{r}}(z_j^k)$ is the character of the representation $\bf{r}$. Explicitly, for a matrix representation, $\rho$, one has 
\begin{align}
    \frac{1}{\mathrm{det}(1-t\rho(g))}&=\mathrm{exp}\left(\sum_{k\geq 1}\frac{t^k \,\mathrm{Tr}\left(\rho(g)^k\right)}{k}\right)\,.
\end{align}}

For Lie groups, the Haar measure has the form 
\begin{align}
    \int_Gd\mu_G &= \frac{1}{(2\pi i)^m}\int_{|z_1|=1}\dots\int_{|z_m|=1}\frac{d z_1}{z_1}\dots\frac{d z_m}{z_m}\prod_{\alpha^{+}}\Bigg( 1-\prod_{i=1}^m z_i^{\alpha^+} \Bigg)
\end{align}
{where $z_i$ are the coordinates of the maximal torus of the gauge group $G$, and $m$ and $\alpha^+$ are the rank and positive roots of $G$, respectively, which exist in the $\bar{\bf{N}}\otimes \bf{N}$ representation of $G$~\cite{Hanany:2008sb}}.

In the 3HDM, as discussed above,  the relevant field content decomposes into four nontrivial
${SU}(3)$ representations: three adjoint representations $\mathbf{8}$ and
one $\mathbf{27}$. Consequently, the Hilbert series 
takes the form of a generating function in four variables, one for each
irreducible component. We denote these variables by $s$, $t$, $u$, and
$q$,
where $s$, $t$, and $u$ correspond to the three $\mathbf{8}$ representations, while
$q$ corresponds to the $\mathbf{27}$. {The grading variables are treated as formal fugacities and, for the contour-integral representation, are taken inside the unit disk.}

{We are interested in polynomial invariants of the 3HDM couplings under
$SU(3)$ basis transformations. The corresponding Hilbert series, which counts
independent invariant polynomials, may be written as~\cite{Bento:2021hyo}}
\begin{align}
    H(s,t,u,q) = & \frac{1}{(2 \pi i)^2} \oint_{|z_1|=1} \frac{d z_1}{z_1}
    \oint_{|z_2|=1} \frac{d z_2}{z_2} (1-z_1 z_2)\left( 1 - \frac{z_1^2}{z_2} \right)
    \left( 1 - \frac{z_2^2}{z_1} \right) \times \nonumber \\[5mm]
    & \mathrm{PE}[z_1,z_2,s,\mathbf{8}] \, \mathrm{PE}[z_1,z_2,t,\mathbf{8}] \,
     \mathrm{PE}[z_1,z_2,u,\mathbf{8}] \, \mathrm{PE}[z_1,z_2,q,\mathbf{27}] \,. 
\end{align}

The functions $\mathrm{PE}[z_1,z_2,s,\mathbf{8}]$ and $\mathrm{PE}[z_1,z_2,q,\mathbf{27}]$ are 
calculated to be~\cite{Bento:2021hyo}
\begin{align}\label{eq:ple8}
    \mathrm{PE}[z_1,z_2,s,\mathbf{8}] = & \left[
    (1-s)^2 \left(1-s \frac{z_1^2}{z_2}\right) \left(1-s \frac{z_2}{z_1^2}\right)
    \left(1-s \frac{1}{z_1 z_2}\right) \times \right.
    \nonumber \\[2mm]
    & \left. \left(1-s z_1 z_2\right) \left(1-s \frac{z_1}{z_2^2}\right) \left(1-s \frac{z_2^2}{z_1}\right) \right]^{-1}
    \, ,
\end{align}
and
\begin{align}\label{eq:ple27}
    \mathrm{PE}[z_1,z_2,q,\mathbf{27}] = & \left[
    (1-q)^3 \left(1-q \frac{1}{z_2^3} \right) \left(1-q z_2^3\right) \left(1-q \frac{1}{z_1 z_2} \right)^2 
    \left(1-q z_1 z_2 \right)^2 \left(1-q \frac{z_2^2}{z_1} \right)^2 \times \right. \nonumber \\[2mm]
    &  \left(1-q \frac{z_1}{z_2^2}\right)^2 \left(1-q \frac{z_2}{z_1^2} \right)^2 \left(1- q \frac{z_1^2}{z_2} \right)^2 
    \left(1- q \frac{1}{z_1^2 z_2^2} \right) \left(1 - q z_1^2 z_2^2 \right)  \times \nonumber \\[2mm]
    &  \left(1-q \frac{z_2^4}{z_1^2} \right) \left(1-q \frac{z_1^2}{z_2^4} \right) 
    \left(1-q \frac{1}{z_1^3} \right) \left(1-q z_1^3\right)
    \left(1-q \frac{z_2^3}{z_1^3} \right)  \times \nonumber \\[2mm]
    & \left.  \left(1-q \frac{z_1^3}{z_2^3} \right) \left(1-q \frac{z_2^2}{z_1^4} \right) 
    \left(1-q \frac{z_1^4}{z_2^2} \right) \right]^{-1} \, .
\end{align}
The ungraded Hilbert series is recovered by identifying all grading variables,
$s=t=u=q$.
This specialization counts invariant operators according only to their total degree and therefore retains less information than the multigraded Hilbert series. Nevertheless, the ungraded Hilbert series remains a useful quantity, as it determines the total number of independent invariants at each order. The corresponding integral was evaluated in Ref.~\cite{Bento:2021hyo}. In addition, Ref.~\cite{Bento:2021hyo} obtained a power-series expansion of the Molien--Weyl integrand about $z_1=z_2=0$, allowing the Hilbert series to be determined through low orders in the couplings. These results provide valuable consistency checks on the multigraded Hilbert series computed in the present work.

\section{3HDM Hilbert Series  Calculation}\label{hseriesCalc}

The evaluation of this contour integral for the Hilbert series however presents several difficulties. {Previous studies of the 3HDM Hilbert series have been limited either to local expansions of the integrand or to the computation of the ungraded Hilbert series~\cite{Bento:2021hyo}.} 

{In this section, we describe these difficulties and the sequence of standard computational techniques used to overcome them. The methods employed are based on established tools from Hilbert-series computations, complex analysis, and invariant theory rather than a new algorithmic framework. Readers interested only in the final form of the Hilbert series may skip directly to Section~\ref{HSResults}.}

\subsection{Issues with the Direct Computation}
A direct evaluation of the integral faces several serious technical obstacles. First, the integrand typically contains higher-order poles arising from quadratic factors in the denominator, such as $(1 - q z_1 z_2)^2$. These lead to poles whose residues require the evaluation of derivatives and rapidly become algebraically cumbersome when many such factors are present. In practice, this already renders a naive residue computation inefficient and unstable in symbolic computation software. 

A second complication arises from the structure of  terms with quadratic or quartic poles in either $z_1$ or $z_2$. Depending on the order of integration, performing one contour integral can introduce branch points in the remaining complex variable, thereby generating branch cuts that obstruct a straightforward application of the residue theorem. This complicates the analytic structure of the integrand and obscures the interpretation of the contour prescription. 

Finally, even if one were to formally classify all poles of the integrand, their number typically grows into the thousands. Summing the corresponding residues is computationally prohibitive in practice, either in \texttt{Mathematica} and in numerical or symbolic implementations in \texttt{Python}~\cite{Bento:2021hyo}. 

\subsubsection{Higher Order Poles}
We first address the higher order poles. These factors
are
\begin{align}
    \bigg[\left(1-q \frac{z_2}{z_1^2} \right)^2 \left(1- q \frac{z_1^2}{z_2} \right)^2 \left(1-q \frac{1}{z_1 z_2} \right)^2 
    \left(1-q z_1 z_2 \right)^2 
       \left(1-q \frac{z_2^2}{z_1} \right)^2 \left(1-q \frac{z_1}{z_2^2}\right)^2\bigg]^{-1}\, .
\end{align}

These higher-order poles can  be handled by rewriting the expression in
terms of derivatives. 
For example, we can write 
\begin{align}
    \bigg[\left(1-q \frac{z_2}{z_1^2} \right)^2\left(1-q \frac{z_1^2}{z_2} \right)^2\bigg]^{-1}\, =
      \frac{\partial^2}{\partial{q_2}\partial{q_1}}\bigg[\left(1-q_1 \frac{z_2}{z_1^2} \right) \left(1-q_2 \frac{z_1^2}{z_2} \right)\bigg]^{-1}\Bigg|_{q_1\to q, q_2\to q }
\end{align}
Since we have six higher order poles, we introduce six variables $q_1,q_2,\dots,q_6$. These are treated as independent during the
differentiation, {a standard technique in complex analysis}. After performing the contour integrals, we take the limit in
which all $q_i$ are set equal to $q$. 

The Hilbert Series can be written as 
\begin{align}
    H(s,t,u,q) =
    %\nonumber\\
    \frac{\partial^6}{\partial{q_1}\partial{q_2}\dots\partial{q_6}}\oint d z_1
     d z_2~ \frac{N(z_1,z_2)}{D(z_1,z_2,s,\dots, q)\times D_{cubic}(z_1,z_2,q)}\Bigg|_{q_1\to q,\dots,q_6\to q }
\end{align}
where 
\begin{align}
    N(z_1,z_2) &=  \frac{1}{(1-s)^2(1-t)^2(1-u)^2(1-q)^3} \frac{1}{(2 \pi i)^2}  \times \frac{(1-z_1 z_2)(1-\frac{z_1^2}{z_2})(1-\frac{z_2^2}{z_1})}{z_1z_2}\,,
\end{align}
and the denominator of the integrand has now only simple poles 
\begin{align}
    D(z_1,z_2,s,\dots,q) =& \prod_{i}^6(1-\alpha^-_i \frac{z_2}{z_1^2})\prod_{i}^6(1-\alpha^+_i \frac{z_1^2}{z_2})\prod_{i}^6(1-\beta^-_i \frac{1}{z_1 z_2})\prod_{i}^6(1-\beta^+_i z_1 z_2)\times \nonumber\\
    &\hspace{15 mm}\prod_{i}^6(1-\gamma^-_i \frac{z^2_2}{z_1})\prod_{i}^6(1-\gamma^+_i \frac{z_1}{z_2^2}) \,,
    \\
   D_{cubic}(z_1,z_2,q)=& \left(1-q \frac{1}{z_2^3} \right) \left(1-q z_2^3\right)
    \left(1-q \frac{1}{z_1^3} \right) \left(1-q z_1^3\right)
    \left(1-q \frac{z_2^3}{z_1^3} \right)  \left(1-q \frac{z_1^3}{z_2^3} \right). 
\end{align}

Here, we have  defined the sets of variables
\begin{align}
    \alpha^- &= \{s,t,u,\sqrt{q},-{\sqrt{q}},q_1\} \label{eq:alpha-minus} \\
    \alpha^+ &= \{s,t,u,\sqrt{q},-{\sqrt{q}},q_2\} \label{eq:alpha-plus} \\
    \beta^-  &= \{s,t,u,\sqrt{q},-{\sqrt{q}},q_3\} \label{eq:beta-minus} \\
    \beta^+  &= \{s,t,u,\sqrt{q},-{\sqrt{q}},q_4\} \label{eq:beta-plus} \\
    \gamma^- &= \{s,t,u,\sqrt{q},-{\sqrt{q}},q_5\} \label{eq:gamma-minus} \\
    \gamma^+ &= \{s,t,u,\sqrt{q},-{\sqrt{q}},q_6\}\,. \label{eq:gamma-plus}
\end{align}

Since we are assuming all  variables $s,t,u,q$ have magnitude $<1$, 
variables in the  $\alpha^-$ array are the ones which contribute poles during the $z_1$ integration while variables in the $\alpha^+$ array do not contribute poles.

\subsubsection{The Partial Fraction Expansion}

Despite the first simplification, the integrand still has too many poles to be evaluated by symbolic codes. We therefore continue the simplification.

We decompose the following functions into partial fraction form
\begin{align}
    A &= \bigg[\prod_{i}^6(1-\alpha^-_i \frac{z_2}{z_1^2})\prod_{i}^6(1-\alpha^+_i \frac{z_1^2}{z_2})\bigg]^{-1}&= \sum_{i=1}^{6} A^-_i\frac{z_2}{z_1-\alpha_i^-z_2^2 }+\sum_{i=1}^{6} A^+_i\frac{z_2}{\alpha_i^+z_1-z_2^2 }\\
    B &= \bigg[\prod_{i}^6(1-\beta^-_i \frac{1}{z_1 z_2})\prod_{i}^6(1-\beta^+_i z_1 z_2)\bigg]^{-1}&= \sum_{i=1}^{6} B^-_i\frac{1}{z_1 z_2-\beta_i^- }+\sum_{i=1}^{6} B^+_i\frac{1}{\beta_i^+z_1 z_2- 1}\\
    C &= \bigg[\prod_{i}^6(1-\gamma^-_i \frac{z^2_2}{z_1})\prod_{i}^6(1-\gamma^+_i \frac{z_1}{z_2^2})\bigg]^{-1} &= \sum_{i=1}^{6} C^-_i\frac{z_2^2}{z_1^2-\gamma_i^-z_2 }+\sum_{i=1}^{6} C^+_i\frac{z_2^2}{\gamma_i^+z_1^2-z_2 }
\end{align}
where
\begin{align}
    A_i^-&= \frac{(\alpha_i^-)^6 }{\prod_{j}(1-\alpha_i^- \alpha_j^+ )\prod_{j\neq i}(\alpha_i^- - \alpha_j^- )}\\
    A_i^+&= \frac{(\alpha_i^+)^5 }{\prod_{j}(1-\alpha_i^+ \alpha_j^- )\prod_{j\neq i}(\alpha_i^+ - \alpha_j^+ )}
\end{align}
and $B^{-}_{i}$, $B^{+}_{i}$, $C^{-}_{i}$, and $C^{+}_{i}$ can similarly be  computed.

This transformation reduces the Hilbert series integrand from an expression that is
prohibitively difficult to evaluate to one that is considerably more tractable.
Naively, the computation requires the evaluation and summation of
$6\times6\times6\times8 = 1728$ complex contour integrals.
However, this number can be reduced  by
temporarily treating the variables $\alpha_i^{\pm}$, $\beta_i^{\pm}$, and
$\gamma_i^{\pm}$ as independent dummy variables, which we denote by
$a$, $b$, and $c$, respectively.

We then need to calculate eight distinct integrals 
\begin{align}
   I_1(a,b,c)=& \oint_{|z_1|=1} d z_1
    \oint_{|z_2|=1} d z_2 \frac{(z_2)^3}{(z_1^2-a z_2)(z_1 z_2-b)(z_1 - c z_2^2)}\frac{N(z_1,z_2)}{D_{cubic}(z_1,z_2)}\\
    I_2(a,b,c)=&\oint_{|z_1|=1} d z_1
    \oint_{|z_2|=1} d z_2 \frac{(z_2)^3}{(a z_1^2- z_2)(z_1 z_2-b)(z_1 -  cz_2^2)}\frac{N(z_1,z_2)}{D_{cubic}(z_1,z_2)}\\
    I_3(a,b,c)=&\oint_{|z_1|=1} d z_1
    \oint_{|z_2|=1} d z_2 \frac{(z_2)^3}{(z_1^2-a z_2)(bz_1 z_2-1)(z_1 - c z_2^2)}\frac{N(z_1,z_2)}{D_{cubic}(z_1,z_2)}\\
    \vdots\nonumber\\
    I_8(a,b,c)=&\oint_{|z_1|=1} d z_1
    \oint_{|z_2|=1} d z_2 \frac{(z_2)^3}{( a z_1^2- z_2)(b z_1 z_2-1)(c z_1 -  z_2^2)}\frac{N(z_1,z_2)}{D_{cubic}(z_1,z_2)}\,.
\end{align}

\subsubsection{Calculation of Residues}

Each of these integrands has nine distinct factors and can now be calculated within the computational powers of symbolic programs. However, there is still a technical issue related to the occurrence of poles with square roots of the second variable.

To see how this problem arises, consider
the factors in the  denominator of the integrand in $I_1$ 
\begin{align}
    & [ (z_1^2-a z_2)(z_1 - c z_2^2)]^{-1}\,.
\end{align}
If the $z_1$
contour integral were performed first, these poles would occur at
$z_1=\pm\sqrt{a z_2}$, and would lead to an integral depending not just on $z_2$ but on its square root.
This would introduce an apparent branch cut
in the subsequent $z_2$ integration.

To avoid this issue, we first perform a partial fraction decomposition of the
integrand with respect to the variable $z_1$, treating $z_2$ as a fixed
parameter. This isolates the factor $(z_1^2-a z_2)$ into a separate term. For the
term containing this factor, we then evaluate the $z_2$ contour integral before
carrying out the $z_1$ integration. With this approach, the integration over $z_2$ and the subsequent $z_1$ integral can be evaluated using
standard residue techniques without introducing any branch-cut ambiguities.

A second issue is  the appearance of poles involving
ratios of the parameters $a,b,c,$ and $q$. For example, 
after performing the first contour integration 
over $z_1$, 
we may obtain new terms in the integrand of the form, for example, $(b^2-a z_2^3)$ with  poles proportional to expressions such as  $b^2/a$. Ambiguities arise for such poles involving ratios of variables, such as
$s/q$ or $q/s$, and  even expressions like $q/q = 1$. Since no
relative ordering between the variables is assumed, it is not a priori clear
whether such poles should be enclosed by the integration contour. 

Fortunately, we find, both analytically and computationally, that all poles involving ratios
of variables cancel in the full expression for the integrand. As a result, the
final Hilbert series is independent of any arbitrary ordering among $s,t,u,$ and
$q$. 

We can therefore perform the contour integrals over $z_1, z_2$ for each $I_i$  using
\texttt{Mathematica}.

\subsection{Calculation of Hilbert Series}
Once the integrals $I_1,\dots,I_8$ have been evaluated, the Hilbert series can be
written as
\begin{align}
    H(s,t,u,q)
    = 
      \sum_{i=1}^6\sum_{j=1}^6\sum_{k=1}^6 I_{ijk}\,,
\end{align}
where the quantities $I_{ijk}$ are 
\begin{align}
I_{ijk} =&\;
\frac{\partial^6}{\partial q_1\,\partial q_2\cdots\partial q_6}
\Big(
A_i^- B_j^- C_k^-\, I_1(\alpha_i^-,\beta_j^-,\gamma_k^-)
+\cdots \nonumber\\
&\hspace{45mm}
+ A_i^+ B_j^+ C_k^+\, I_8(\alpha_i^+,\beta_j^+,\gamma_k^+)
\Big)
\Bigg|_{q_1=\cdots=q_6=q}\,.
\end{align}

For fixed values of $i$, $j$, and $k$, the computation of each individual
$I_{ijk}$ is straightforward.  Now, however, each $I_{ijk}$ is of the form $I_{ijk}=N_{ijk}/D_{ijk}$, where both the numerator and denominator are
lengthy expressions. For example, the denominator of $I_{6,6,6}$ contains on the
order of one hundred factors, while its expanded numerator contains roughly
$10^4$ terms, with coefficients as large as $\mathcal{O}(10^{10})$ and monomials
of degree up to $\mathcal{O}(s^7 t^7 u^7 q^{59})$. 
This complexity makes  summation  over $i,j,k$ computationally difficult.

To address this issue, we  first bring the sum of terms to a common denominator
\begin{align}
    \sum_{i,j,k} I_{ijk}
    = \frac{N_{\mathrm{LCM}}}{D_{\mathrm{LCM}}}\,.
\end{align}

To compute $D_{{LCM}}$  we used
the \texttt{PolynomialLCM} function in \texttt{Mathematica}. Computing $N_{{LCM}}$, however,
poses a significantly greater challenge. Indeed, one may write
\begin{align}
    N_{LCM}
    =& \sum_{ijk}N_{ijk}\times\frac{ D_{LCM}}{D_{ijk}}\\
   = & \sum_{ijk}N_{ijk}\times{ D'_{ijk}}
\end{align}
where $D'_{ijk}$ is obtained by canceling all factors of $D_{ijk}$ from
$D_{\mathrm{LCM}}$, which is always possible since $D_{\mathrm{LCM}}$ is the least
common multiple of all $D_{ijk}$.

Fully expanding each product $N_{ijk} \times D'_{ijk}$ is computationally
demanding. The expressions $D'_{ijk}$ typically contain hundreds of factors, the
maximum powers of $s$, $t$, $u$, and $q$ appearing in
$N_{ijk} \times D'_{ijk}$ are of order $90$, $90$, $90$, and $300$, respectively,
and the integer coefficients in these expansions can reach values as large as
$\mathcal{O}(10^{27})$. Furthermore, at this stage $N_{LCM}$ and $D_{LCM}$ are not polynomials in $q$, but rather in $\sqrt{q}$, due to the dependence introduced by the arrays defined in Eqs.~\eqref{eq:alpha-minus}--\eqref{eq:gamma-plus}.

We found that the most effective way to address the computational complexity is
to stop using symbolic algebra packages such as \texttt{Mathematica} or
\texttt{Python}'s \texttt{SymPy}, and instead work directly with coefficient
lists of \texttt{Python} integers. Since the coefficients can exceed
$2^{63}-1 \simeq 9\times 10^{18}$, standard \texttt{NumPy} \texttt{Int64} types
are insufficient, and one must instead use arbitrary-precision \texttt{Python}
integers. 

Defining $Q \equiv \sqrt{q}$, we write 
\begin{align}
    N_{ijk}=& \sum_{i_1,i_2,i_3,i_4} A_{i_1,i_2,i_3,i_4} s^{i_1} t^{i_2}u^{i_3}Q^{i_4}
\end{align}
where the coefficient tensor $A_{i_1,i_2,i_3,i_4}$ can be computed directly in
\texttt{Mathematica} using the \texttt{CoefficientList} function. 

Now, for instance, to multiply $N_{ijk}$ by the first factor appearing in
$D'_{ijk}$—such as $(-1+q)$—we compute
\begin{align*}
    N_{ijk}(-1+q)
    =& -\left(\sum_{i_1,i_2,i_3,i_4} A_{i_1,i_2,i_3,i_4} s^{i_1} t^{i_2}u^{i_3}Q^{i_4}\right) +\left(\sum_{i_1,i_2,i_3,i_4} A_{i_1,i_2,i_3,i_4} s^{i_1} t^{i_2}u^{i_3}Q^{i_4+2}\right)\,.
\end{align*}
 This multiplication then  reduces to summing two four-dimensional arrays. This procedure
generalizes to all factors appearing in $D'_{ijk}$.

Since each product $N_{ijk} D'_{ijk} \equiv N'_{ijk}$ is a four-dimensional array
with dimensions as large as $(90,90,90,600)$—corresponding to the maximal
exponents of $s$, $t$, $u$, and $\sqrt{q}$—and with integer coefficients reaching
values of order $10^{27}$, this procedure consumes a substantial amount of RAM.
Nevertheless, it is still computationally less demanding than manipulating the
corresponding polynomials directly in \texttt{Mathematica} or \texttt{Python}'s
\texttt{SymPy}. 

At this stage, the Hilbert series has been written in the form
\begin{align}
     H(s,t,u,q)&=  \frac{N_{LCM}}{D_{LCM}}\,,
 \end{align}
 where
$D_{\mathrm{LCM}}$ is the least common multiple of the denominators appearing in
the intermediate expressions. 
However, $N_{\mathrm{LCM}}$ and $D_{\mathrm{LCM}}$ may still contain
common polynomial factors. 

Dividing out all common factors using standard polynomial algorithms in
\texttt{Mathematica} or \texttt{Python} would require writing
$N_{\mathrm{LCM}}$ explicitly as a polynomial, which is prohibitively memory
intensive. Instead, it is far more practical to retain
$N_{\mathrm{LCM}}$ in its coefficient-array representation and carry out the
division directly at the level of arrays polynomial synthetic division formulas.

For example, consider the factor, $(s-t)$.  We first verify whether this divides $N_{{LCM}}
$ by checking whether $N_{{LCM}}=0$ when $s=t$.
This can be done straightforwardly at the level of the coefficient array for
$N_{{LCM}}$. 
If $N_{LCM}$ is divisible by $(s-t)$, we write 
\bea
N_{LCM}=\sum a_i(t,u,Q)s^i\qquad {N_{LCM}\over (s-t)}=\sum b_i(t,u,Q)s^i
\eea
One derives 
\begin{align}
    b_{n-1}= a_n\qquad
    b_i= a_{i+1}+ t b_{i+1}
\end{align}
where each $a_{i}$ and $b_i$ is a three dimensional coefficient array in $t$, $u$, and $q$. 
The division can then be done directly by manipulating the coefficient arrays.

Using these methods, we can find an explicit expression for the Hilbert series after all common factors have been divided through.

\section{Results for the 3HDM Hilbert Series }\label{HSResults}
In this section, we present the results of our Hilbert series calculation for the
3HDM. {The computation of the multigraded Hilbert series was carried out using several independent Mathematica and Python scripts, including the evaluation of the Molien--Weyl integral, partial-fraction decomposition, and the subsequent synthetic division used to obtain the minimal form of the Hilbert series. As these calculations were performed in multiple stages, an exact total runtime is difficult to determine. However, the cumulative runtime was approximately 100 CPU-hours.}

{The calculation required access to a high-memory computing cluster with 124 GB of RAM. In particular, several intermediate expressions exceeded the memory available on standard cloud-computing instances with 51 GB of RAM, preventing the full calculation from being completed in that environment.}

The Hilbert series can be written as 
\begin{align}
     H(s,t,u,q)&=  \frac{N}{D}
 \end{align}
 where the denominator is
 \begin{align}
D = {}&(1-q)(1+q)^2(1-q^3)^5(1-q^4)^4(1-q^5)^4(1-q^6)^2(1-q^7)^2(1-q^9)\nonumber\\
&(1-qs)(1-q^2 s)^3(1-q^3 s)^2(1-s^2)(1-q s^2)^4(1-q^2 s^2)^2(1-q^3 s^2)^2\nonumber\\
&(1-s^3)(1-q s^3)^2(1-q^2 s^3)(1-q^3 s^3)(1-q s^4)(1-q^3 s^4)\nonumber\\
&(1-q t)(1-q^2 t)^3(1-q^3 t)^2(1-s t)^2(1-q s t)^2(1-s^2 t)(1-q s^2 t)\nonumber\\
&(1-t^2)(1-q t^2)^4(1-q^2 t^2)^2(1-q^3 t^2)^2(1-s t^2)(1-q s t^2)(1-q s^2 t^2)\nonumber\\
&(1-t^3)(1-q t^3)^2(1-q^2 t^3)(1-q^3 t^3)(1-q t^4)(1-q^3 t^4)\nonumber\\
&(1-q u)(1-q^2 u)^3(1-q^3 u)^2(1-s u)^2(1-q s u)^2(1-s^2 u)(1-q s^2 u)\nonumber\\
&(1-t u)^2(1-q t u)^2(1-s t u)(1-t^2 u)(1-q t^2 u)\nonumber\\
&(1-u^2)(1-q u^2)^4(1-q^2 u^2)^2(1-q^3 u^2)^2(1-s u^2)(1-q s u^2)(1-q s^2 u^2)\nonumber\\
&(1-t u^2)(1-q t u^2)(1-q t^2 u^2)(1-u^3)(1-q u^3)^2(1-q^2 u^3)(1-q^3 u^3)\nonumber\\
&(1-q u^4)(1-q^3 u^4)\,.
\end{align}
 
 The fully expanded numerator of the multigraded Hilbert series contains 31,689,364 terms and is therefore far too large to display explicitly.  The complete numerator will therefore be made available in an online repository~\cite{Bryan:HSdata}.
Here, we present the first few   terms of the numerator.  
Up to $\mathcal{O}\big((s t u q)^6\big)$, the numerator is 
\begin{align}
    N = {} & 1
+ s^{3}tu + s^{2}t^{2}u + s^{2}t^{2} + s^{2}tu^{2} + 3s^{2}tu + s^{2}u^{2} + st^{3}u + st^{2}u^{2} + 3st^{2}u
+ stu^{3} \nonumber\\
&+ 3stu^{2} + stu - st - su + t^{2}u^{2} - tu + q - qu - 3qu^{2} - qu^{3} - qt - 2qtu \nonumber\\
&+ 3qtu^{2} + 7qtu^{3} - 3qt^{2} + 3qt^{2}u + 8qt^{2}u^{2} - qt^{3} + 7qt^{3}u - qs - 2qsu + 3qsu^{2} \nonumber\\
&+ 7qsu^{3} - 2qst + 10qstu + 20qstu^{2} + 3qst^{2} + 20qst^{2}u + 7qst^{3} - 3qs^{2} + 3qs^{2}u \nonumber\\
&+ 8qs^{2}u^{2} + 3qs^{2}t + 20qs^{2}tu + 8qs^{2}t^{2} - qs^{3} + 7qs^{3}u + 7qs^{3}t - 3q^{2}u - 2q^{2}u^{2} \nonumber\\
&+ 4q^{2}u^{3} - 3q^{2}t + 4q^{2}tu + 20q^{2}tu^{2} - 2q^{2}t^{2} + 20q^{2}t^{2}u + 4q^{2}t^{3} - 3q^{2}s + 4q^{2}su \nonumber\\
&+ 20q^{2}su^{2} + 4q^{2}st + 40q^{2}stu + 20q^{2}st^{2} - 2q^{2}s^{2} + 20q^{2}s^{2}u + 20q^{2}s^{2}t \nonumber\\
&+ 4q^{2}s^{3} - 3q^{3}u + 9q^{3}u^{2} - 2q^{3}t + 26q^{3}tu + 9q^{3}t^{2} - 2q^{3}s + 26q^{3}su \nonumber\\
&+ 26q^{3}st + 9q^{3}s^{2} - 3q^{3} - 4q^{4} + 10q^{4}u + 10q^{4}t + 10q^{4}s - q^{5}+\mathcal{O}\big((s t u q)^6\big)\,.
\end{align}

The Hilbert series can be expanded as a power series, and we find that this expansion exactly reproduces the results of~\cite{Bento:2021hyo}. In addition, ungraded Hilbert series may be obtained by setting appropriate subsets of the grading variables equal to one another or to zero. Performing this procedure provides a further consistency check, and we verify that the resulting ungraded Hilbert series also agrees with the results of~\cite{Bento:2021hyo}.

\section{Calculation of Invariants up to $\mathcal{O}(\lambda^3)$}\label{Invariants}
We now return to the 3HDM (discussed above in Section~\ref{3hdmoverview}) and attempt to construct the invariants of this theory. As described there, physical observables must be polynomials in the parameters $(\mu^2)_{i}^{~j}$ and $\lambda_{ij}^{~~kl}$ that are invariant under the $SU(3)$ group of field redefinitions. Consequently, the problem of identifying physical observables reduces to determining all polynomial combinations of these parameters that remain invariant under this symmetry. 

Recall, under the  $SU(3)$ action, the mass parameters $(\mu^2)_{i}^{~j}$
transform in the representation 
\begin{align}
    \bar{\mathbf{3}} \otimes \mathbf{3}
    &= \mathbf{1} \oplus \mathbf{8}\,,
\end{align}
Explicitly, these are the $SU(3)$ singlet 
 \begin{align}
 det(\mu^2)=\mu^2_{11}\mu^2_{22}\mu^2_{33}
 \end{align}
 as well as an $SU(3)$ adjoint
\begin{align}
    (V^1)_i^{~j} &= (\mu^2)_{i}^{~j}-\frac{1}{3}\sum_k(\mu^2)_k^{~k}\delta_i^{~j}\,.
\end{align}

$\lambda_{ij}^{~~kl}$ 
{contains two adjoints; these 
are}
\begin{align}
(V^{(2)})_i^{~k}=& \sum_{l,j}\left(\lambda_{ij}^{~~kl}\delta_l^{~j}+\lambda_{ij}^{~~lk}\delta_l^{~j}\right)-\frac{1}{3}\sum_{n,m}(\lambda_{nm}^{~~nm}+\lambda_{nm}^{~~mn})\delta_i^{~k}
\\
(V^{(3)})_m^{~n}=& \sum_{i,j,k,l}\lambda_{ij}^{~~kl}\epsilon_{klm}\epsilon^{ijn}-\frac{1}{3}\sum_{i,j,k,l,p}\lambda_{ij}^{~~kl}\epsilon_{klp}\epsilon^{ijp}\delta_m^{~n}
\end{align}
{and the remaining components are organized into a traceless $W_{ij}^{~~kl}$ in the $\bf{27}$ of $SU(3)$,}
\begin{multline}
    W_{ij}^{~~kl} =\lambda_{ij}^{~~kl}+\lambda_{ji}^{~~kl}+\frac{1}{20}\sum_{m,n}(\lambda_{mn}^{~~mn}+\lambda_{mn}^{~~nm})(\delta_{i}^{~k}\delta_{j}^{~l}+\delta_{j}^{~k}\delta_{i}^{~l})\\-\frac{1}{5}\sum_{n}\big((\lambda_{in}^{~~nk}+\lambda_{in}^{~~kn})\delta_{j}^{~l}+(\lambda_{jn}^{~~nk}+\lambda_{jn}^{~~kn})\delta_{i}^{~l}\\+(\lambda_{in}^{~~nl}+\lambda_{in}^{~~ln})\delta_{j}^{~k}+(\lambda_{jn}^{~~nl}+\lambda_{jn}^{~~ln})\delta_{i}^{~k}\big)\,.
\end{multline}

{We now look for combinations of these parameters which are invariant under the symmetry. The number of independent invariants grows rapidly with the degree of the operator. We therefore restrict our explicit classification to invariants containing up to three quartic couplings, $\mathcal{O}(\lambda^3)$. This already yields a substantial set of independent invariants and is sufficient to illustrate the structure of the invariant ring. (For the special case of invariants containing no $\mathbf{27}$ fields, we extend the analysis to arbitrary order in the quartic coupling $\lambda$.) A complete classification of the full invariant basis is left for future work.}

Unlike the quartic couplings, however, we do not impose any restriction on the number of mass insertions. Consequently, the invariants may contain arbitrary powers of the mass parameters.
In our case, the masses are in a $\bf{1}\otimes \bf{8}$ of $SU(3)$. The $\bf{1}$ is an invariant by itself, and can be factored out of any invariants in which it occurs;  however, the $\bf{8}$ cannot be treated so easily. Invariants may a priori contain several factors of the $\bf{8}$.

To find these invariants, we use a method motivated by the background field approaches in field theory~\cite{Luty:1995sd,Brax:2001an}.

\subsection{The Background Field Approach}
We move to the mass basis, where   the mass terms are diagonal
\begin{align}~\label{bkrnd}
    \mu^2 &= 
    \begin{pmatrix}
        \mu^2_{11}&0&0 \\
        0&\mu^2_{22}&0 \\
        0&0&\mu^2_{33}
    \end{pmatrix}.
\end{align}
{We assume a generic point in parameter space with non-degenerate masses. This may be viewed as choosing a representative on a generic $SU(3)$ orbit, or equivalently as turning on background values for the singlet and adjoint components of the mass matrix.}

{For a generic diagonal mass matrix, the subgroup of $SU(3)$ which leaves the background invariant is the maximal torus $U(1)_1\times U(1)_2$, generated by}
\begin{align}
    t^3 = 
    \begin{pmatrix}
        0&0&0 \\
        0&1&0 \\
        0&0&-1
    \end{pmatrix}
    &&
    t^8 = 
    \begin{pmatrix}
        -2&0&0 \\
        0&1&0 \\
        0&0&1
    \end{pmatrix}\,.
\end{align}
We denote the  abelian symmetries generated by $t^3$ and $t^8$ as
$U(1)_1$ and $U(1)_2$, respectively. 

 The invariants of the theory, which were originally polynomials in  $(\mu^2)_{i}^{~j}$ and $\lambda_{ij}^{~~kl}$ invariant under $SU(3)$, now just become   polynomials in  $\lambda_{ij}^{~~kl}$ which are invariant under the residual $U(1)_1\times U(1)_2$ symmetry. We shall refer to these below as the $U(1)_1\times U(1)_2$ invariants.

\subsection{The $U(1)_1\times U(1)_2$ Invariants}\label{u1u1invariants}
To find these invariants, we first find the charges of the components of  $\lambda_{ij}^{~~kl}$.
These are summarized 
in the table
below. The first column lists all distinct $U(1)_1 \times U(1)_2$ charges, while
the second column indicates the corresponding field components that carry each
charge. 

In each case, we have eliminated components which are constrained. For example, since the adjoint is traceless, we must have
$(V^I)_{1}^{~1}+(V^I)_{2}^{~2}+(V^I)_{3}^{~3}=0$, we can eliminate
$(V^I)_{3}^{~3}$  in favor of the other fields. Similarly, we can eliminate components of $W$ with both a fundamental 3 index as well as an antifundamental 3 index.

\begin{center}
\renewcommand{\arraystretch}{1.10}
    \begin{tabular}{|c|c|}
        \hline
        $U(1)_1\times U(1)_2$ Charge &Field(s) \\
        \hline
        $Q_{4,0}$ & $W_{22}^{~~33}$\\
        \hline
        $Q_{3,1}$ & $W_{22}^{~~13}$\\
        \hline
        $Q_{3,-1}$ & $W_{12}^{~~33}$\\
        \hline
        $Q_{2,2}$ & $W_{22}^{~~11}$\\
        \hline
        $Q_{2,0}$ & $W_{12}^{~~13}$, $W_{22}^{~~23}$, $(V^2)_{2}^{~3}$, $(V^3)_{2}^{~3}$\\
        \hline
        $Q_{2,-2}$ & $W_{11}^{~~33}$\\
        \hline
        $Q_{1,1}$ & $W_{12}^{~~11}$, $W_{22}^{~~12}$, $(V^2)_{2}^{~1}$, $(V^3)_{2}^{~1}$\\
        \hline
        $Q_{1,-1}$ & $W_{11}^{~~13}$, $W_{12}^{~~23}$, $(V^2)_{1}^{~3}$, $(V^3)_{1}^{~3}$\\
        \hline
        $Q_{0,2}$ & $W_{23}^{~~11}$\\
        \hline
        $Q_{0,0}$ & $W_{11}^{~~11}$, $W_{12}^{~~12}$, $W_{22}^{~~22}$, $(V^2)_{1}^{~1}$, $(V^3)_{1}^{~1}$,$(V^2)_{2}^{~2}$,$(V^3)_{2}^{~2}$\\
        \hline
        $Q_{0,-2}$ & $W_{11}^{~~23}$\\
        \hline
        $Q_{-1,1}$ & $W_{13}^{~~11}$, $W_{23}^{~~12}$, $(V^2)_{3}^{~1}$, $(V^3)_{3}^{~1}$\\
        \hline
        $Q_{-1,-1}$ & $W_{11}^{~~12}$, $W_{12}^{~~22}$, $(V^2)_{1}^{~2}$, $(V^3)_{1}^{~2}$\\
        \hline
        $Q_{-2,2}$ & $W_{33}^{~~11}$\\
        \hline
        $Q_{-2,0}$ & $W_{13}^{~~12}$, $W_{23}^{~~22}$, $(V^2)_{3}^{~2}$, $(V^3)_{3}^{~2}$\\
        \hline
        $Q_{-2,-2}$ & $W_{11}^{~~22}$\\
        \hline
        $Q_{-3,1}$ & $W_{33}^{~~12}$\\
        \hline
        $Q_{-3,-1}$ & $W_{13}^{~~22}$\\
        \hline
        $Q_{-4,0}$ & $W_{33}^{~~22}$\\
        \hline
    \end{tabular}
\end{center}

It is now straightforward to find combinations of these parameters which are invariant under the residual symmetry. These  $U(1)_1\times U(1)_2$-invariant polynomials are listed in the Appendix.

\subsection{3HDM Matching}\label{3hdmmatching}

The $U(1)_1\times U(1)_2$-invariants  are in fact a complete set of invariants of the 3HDM. However, they are written in a form
that is basis dependent, since we have treated the mass terms as a background field breaking the symmetry. We will now find the invariants with full $SU(3)$ symmetry, which is done by combining the $U(1)_1\times U(1)_2$ invariant polynomials listed in the Appendix with powers of the masses.

As a first  illustrative example, consider the components $(V^2)_1^{~1}$ and $(V^2)_2^{~2}$, which are completely invariant under $U(1)_1\times U(1)_2$. These objects correspond to $\mathcal{O}(\lambda)$ invariants of the theory. 
They must correspond to
${SU}(3)$-invariant contractions involving a single $V^2$  multiplying  an arbitrary number of $V^1$ insertions (recall that we have denoted the adjoint in the mass terms as $V^1$). 

A natural guess for such invariants are
\bea
    I^{1,2} &= \mathrm{Tr}(V^{1}V^{2})\\
    I^{1,1,2} &= \mathrm{Tr}(V^{1}V^{1}V^{2})
\eea

In fact, these two structures are sufficient. Explicitly, if one now treats the mass terms as being the background value, Eqn.~\ref{bkrnd}, then one finds
\begin{align}
    (V^2)_{1}^{~1}&= \frac{2\mu^2_{11}-\mu^2_{22}-\mu^2_{33}}{6(\mu^2_{11}-\mu^2_{22})(\mu^2_{11}-\mu^2_{33})}I^{2,1}_1+\frac{2}{3(\mu^2_{11}-\mu^2_{22})(\mu^2_{11}-\mu^2_{33})}I^{2,1,1}_2\\
    (V^2)_{2}^{~2}&= \frac{\mu^2_{11}-2\mu^2_{22}+\mu^2_{33}}{6(\mu^2_{11}-\mu^2_{22})(\mu^2_{11}-\mu^2_{33})}I^{2,1}_1-\frac{2}{3(\mu^2_{11}-\mu^2_{22})(\mu^2_{11}-\mu^2_{33})}I^{2,1,1}_2\,.
\end{align}

Hence the two invariants $(V^2)_1^{~1}$ and $(V^2)_2^{~2}$ arise from the two invariants $ I^{1,2},
    I^{1,1,2}$. It will turn out to be convenient to keep the flavor structure more general. That is, we will take the invariants to be 
\begin{align}
    I^{I,J} &= \mathrm{Tr}(V^{I}V^{J})\\
    I^{I,J,K} &= \mathrm{Tr}(V^{I}V^{J}V^{K})
\end{align}
where $I,J,K=1,2,3$.

We can extend this procedure to invariants involving the $\mathbf{27}$ representation.
As an example, consider the components $W_{11}^{~~11}$, $W_{12}^{~~12}$, and $W_{22}^{~~22}$, which are completely invariant under the residual $U(1)_1\times U(1)_2$ symmetry. 
They must correspond to
${SU}(3)$-invariant contractions involving a single $W$  multiplying  an arbitrary number of $V^1$ insertions

We can  write several candidate ${SU}(3)$ invariants involving
a single $W$ and up to four factors of $V^1$:
\begin{align}
    I_1&=W_{kl}^{~~ij}(V^{1})_{i}^{~k}(V^{1})_{j}^{~l}\\
    I_2&=W_{kl}^{~~ij}(V^{1}V^1)_{i}^{~k}(V^{1})_{j}^{~l}\\
    I_3&=W_{kl}^{~~ij}(V^{1}V^1)_{i}^{~k}(V^1V^{1})_{j}^{~l}\\
    I_4&=W_{kl}^{~~ij}(V^{1})_{i}^{~k}(V^1V^1V^{1})_{j}^{~l}
\end{align}
Since there are only three independent $U(1)_1\times U(1)_2$ invariants in this
sector, one of the above structures must be linearly dependent on the others.

These invariants can be evaluated straightforwardly once $V^1$ is assigned its
background value. For example, one finds
\begin{align}
    I_1&= (\mu^2_{11}-\mu^2_{33})^2W_{11}^{~~11}+2(\mu^2_{11}-\mu^2_{33})(\mu^2_{22}-\mu^2_{33})W_{12}^{~~12}+(\mu^2_{22}-\mu^2_{33})^2W_{22}^{~~22}
\end{align}
Analogous expressions may be obtained for $I_2$, $I_3$, and $I_4$. One then finds
that $I_1$, $I_2$, and $I_3$ are linearly independent, while
\begin{align}
    I_4&=\mathrm{Tr}(V^1V^1) I_1
\end{align}
and is therefore not an independent invariant.

This procedure can be automated efficiently using \texttt{Mathematica}. For each
set of proposed basis invariants, we use \texttt{NullSpace} to determine the
dimension of the space they generate, and \texttt{SubspaceBasis} to extract a
linearly independent set spanning that space. Repeating this process allows one
to systematically construct a basis of independent ${SU}(3)$ invariants
that reproduces all $U(1)_1\times U(1)_2$ invariants. {The results are listed in Section~\ref{invariantresults}}.

\section{Results}~\label{invariantresults}
By applying the method described in Section~\ref{3hdmmatching}, we find all $SU(3)$-invariants that are required to reproduce the list of invariants in the Appendix. This is therefore a full  set of invariants for the 3HDM up to cubic order in the couplings. In the special case of invariants containing no $\mathbf{27}$ fields, the results are obtained to arbitrary order in the quartic coupling $\lambda$.
It is organized by the number of $W$ followed by the number of $V$.
\ytableausetup{boxsize=1.25em}
\begin{center}
\renewcommand{\arraystretch}{1.25}
    \begin{tabular}{|c|c|}
        \hline
         & 3HDM Invariants \\
        \hline
        \multirow{1}{8em}{Zero $\bf{27}$, Two $\bf{8}$}&$\mathrm{Tr}(V^IV^J)$\\
        \hline
        \multirow{2}{8em}{Zero $\bf{27}$, Three $\bf{8}$}&$\mathrm{Tr}(V^IV^JV^K)$\\
        &$\mathrm{Tr}([V^IV^JV^K])$\\
        \hline
        \multirow{2}{8em}{Zero $\bf{27}$, Four $\bf{8}$}&$\mathrm{Tr}([V^IV^J][V^KV^L])$\\
        &$\mathrm{Tr}([V^IV^JV^K]V^L)$\\
        \hline
        \multirow{2}{8em}{Zero $\bf{27}$, Five $\bf{8}$}&$\mathrm{Tr}([V^IV^JV^K]V^LV^M)$\\
        &$\mathrm{Tr}([V^IV^JV^K][V^LV^M])$\\
        \hline
        \multirow{1}{8em}{Zero $\bf{27}$, Six $\bf{8}$}&$\mathrm{Tr}([V^IV^J][V^KV^L][V^MV^N])$\\
        \hline
        \multirow{1}{8em}{One $\bf{27}$, Two $\bf{8}$}  & ${W_{ij}^{~~kl}(V^I)_{k}^{~i}(V^J)_{l}^{~j}}$\\
        \hline
        \multirow{1}{8em}{One $\bf{27}$, Three $\bf{8}$} &${W_{ij}^{~~kl}(V^{I}V^{J})_{k}^{~i}(V^K)_{l}^{~j}}$\\
        \hline
   \multirow{2}{8em}{One $\bf{27}$, Four $\bf{8}$} &${W_{ij}^{~~kl}(V^{I}V^{J}V^K)_{k}^{~i}(V^L)_{l}^{~j}}$\\
     &${W_{ij}^{~~kl}(V^{I}V^{J})_{k}^{~i}(V^KV^L)_{l}^{~j}}$\\
        \hline
        \multirow{2}{8em}{One $\bf{27}$, Five $\bf{8}$} &$W_{ij}^{~~kl}(V^IV^JV^KV^L)_{k}^{~i}(V^M)_{l}^{~j}$\\
     &$W_{ij}^{~~kl}(V^IV^JV^K)_{k}^{~i}(V^LV^M)_{l}^{~j}$\\
        \hline
    \multirow{2}{8em}{One $\bf{27}$, Six $\bf{8}$} & $W_{ij}^{~~kl}(V^IV^JV^KV^L)_{k}^{~i}(V^MV^N)_{l}^{~j}$\\
     & $W_{ij}^{~~kl}(V^IV^JV^KV^LV^M)_{k}^{~i}(V^N)_{l}^{~j}$\\
        \hline
        \multirow{1}{8em}{One $\bf{27}$, Seven $\bf{8}$} &$W_{ij}^{~~kl}(V^IV^JV^KV^LV^M)_{k}^{~i}(V^NV^O)_{l}^{~j}$\\
        \hline
         \multirow{1}{8em}{Two $\bf{27}$, Zero $\bf{8}$} & $W_{ij}^{~~kl}W_{kl}^{~~ij}$\\
        \hline
         \multirow{1}{8em}{Two $\bf{27}$, One $\bf{8}$} &$W_{ij}^{~~ka}W_{bk}^{~~ij}(V^I)_{a}^{~b}$\\
        \hline
        \multirow{2}{8em}{Two $\bf{27}$, Two $\bf{8}$} &$W_{ij}^{~~ka}W_{bk}^{~~ij}(V^IV^J)_{a}^{~b}$\\
        &$W_{ib}^{~~ak}W_{kd}^{~~ci}(V^I)_{a}^{~b}(V^J)_{c}^{~d}$\\
        \hline
         \multirow{4}{8em}{Two $\bf{27}$, Three $\bf{8}$} &$W_{ib}^{~~ak}W_{kd}^{~~ci}(V^IV^J)_{a}^{~b}(V^K)_{c}^{~d}$\\
        &$W_{fb}^{~~ak}W_{kd}^{~~ce}(V^I)_{a}^{~b}(V^J)_{c}^{~d}(V^K)_{e}^{~f}$\\
        &$W_{df}^{~~ak}W_{kb}^{~~ce}(V^I)_{a}^{~b}(V^J)_{c}^{~d}(V^K)_{e}^{~f}$\\
        &$W_{ij}^{~~ka}W_{bk}^{~~ij}(V^IV^JV^K)_{a}^{~b}$\\
        \hline
\end{tabular}
\end{center}
\begin{center}
\renewcommand{\arraystretch}{1.25}
    \begin{tabular}{|c|c|}
        \hline
         &3HDM Invariants \\
        \hline
        \multirow{9}{8em}{Two $\bf{27}$, Four $\bf{8}$} &$W_{fb}^{~~ak}W_{kd}^{~~ce}(V^IV^J)_{a}^{~b}(V^K)_{c}^{~d}(V^L)_{e}^{~f}$\\
        &$W_{fb}^{~~ak}W_{kd}^{~~ce}(V^I)_{a}^{~b}(V^JV^K)_{c}^{~d}(V^L)_{e}^{~f}$\\
        &$W_{fh}^{~~ac}W_{bd}^{~~eg}(V^I)_{a}^{~b}(V^J)_{c}^{~d}(V^K)_{e}^{~f}(V^L)_{g}^{~h}$\\
        &$W_{fb}^{~~ag}W_{hd}^{~~ce}(V^I)_{a}^{~b}(V^J)_{c}^{~d}(V^K)_{e}^{~f}(V^L)_{g}^{~h}$\\
        &$W_{ij}^{~~ka}W_{bk}^{~~ij}(V^IV^JV^KV^L)_{a}^{~b}$\\
        &$W_{ib}^{~~ak}W_{kd}^{~~ci}(V^IV^JV^K)_{a}^{~b}(V^L)_{c}^{~d}$\\
        &$W_{fb}^{~~ak}W_{kd}^{~~ce}(V^I)_{a}^{~b}(V^J)_{c}^{~d}(V^KV^L)_{e}^{~f}$\\
        &$W_{db}^{~~ek}W_{kf}^{~~ac}(V^I)_{a}^{~b}(V^JV^K)_{c}^{~d}(V^L)_{e}^{~f}$\\
        &$W_{db}^{~~ek}W_{kf}^{~~ac}(V^I)_{a}^{~b}(V^J)_{c}^{~d}(V^KV^L)_{e}^{~f}$\\
        \hline
        \multirow{7}{8em}{Two $\bf{27}$, Five $\bf{8}$} &$W_{fb}^{~~ak}W_{kd}^{~~ce}(V^IV^J)_{a}^{~b}(V^KV^L)_{c}^{~d}(V^M)_{e}^{~f}$\\
        &$W_{fb}^{~~ak}W_{kd}^{~~ce}(V^IV^J)_{a}^{~b}(V^K)_{c}^{~d}(V^LV^M)_{e}^{~f}$\\
        &$W_{fb}^{~~ak}W_{kd}^{~~ce}(V^I)_{a}^{~b}(V^JV^K)_{c}^{~d}(V^LV^M)_{e}^{~f}$\\
        &$W_{df}^{~~ak}W_{kb}^{~~ec}(V^IV^J)_{a}^{~b}(V^KV^L)_{c}^{~d}(V^M)_{e}^{~f}$\\
        &$W_{ib}^{~~ak}W_{kd}^{~~ci}(V^IV^JV^K)_{a}^{~b}(V^LV^M)_{c}^{~d}$\\
        &$W_{fb}^{~~ak}W_{kd}^{~~ce}(V^IV^JV^K)_{a}^{~b}(V^L)_{c}^{~d}(V^M)_{e}^{~f}$\\
        &$W_{fb}^{~~ak}W_{kd}^{~~ce}(V^I)_{a}^{~b}(V^JV^KV^L)_{c}^{~d}(V^M)_{e}^{~f}$\\
        \hline
        \multirow{9}{8em}{Two $\bf{27}$, Six $\bf{8}$ }&$W_{fb}^{~~ak}W_{kd}^{~~ce}(V^IV^J)_{a}^{~b}(V^KV^L)_{c}^{~d}(V^MV^N)_{e}^{~f}$\\
        &$W_{fh}^{~~ac}W_{bd}^{~~eg}(V^IV^J)_{a}^{~b}(V^KV^L)_{c}^{~d}(V^M)_{e}^{~f}(V^N)_{g}^{~h}$\\
        &$W_{fb}^{~~ag}W_{hd}^{~~ce}(V^IV^J)_{a}^{~b}(V^K)_{c}^{~d}(V^LV^M)_{e}^{~f}(V^N)_{g}^{~h}$\\
        &$W_{fb}^{~~ak}W_{kd}^{~~ce}(V^IV^JV^K)_{a}^{~b}(V^L)_{c}^{~d}(V^MV^N)_{e}^{~f}$\\
        &$W_{fb}^{~~ak}W_{kd}^{~~ce}(V^IV^JV^K)_{a}^{~b}(V^LV^M)_{c}^{~d}(V^N)_{e}^{~f}$\\
        &$W_{fb}^{~~ak}W_{kd}^{~~ce}(V^IV^J)_{a}^{~b}(V^KV^LV^M)_{c}^{~d}(V^N)_{e}^{~f}$\\
        &$W_{fb}^{~~ak}W_{kd}^{~~ce}(V^I)_{a}^{~b}(V^JV^KV^L)_{c}^{~d}(V^MV^N)_{e}^{~f}$\\
        &$W_{bd}^{~~ek}W_{kf}^{~~ac}(V^IV^J)_{a}^{~b}(V^KV^L)_{c}^{~d}(V^MV^N)_{e}^{~f}$\\
         &$W_{fb}^{~~ag}W_{hd}^{~~ce}(V^IV^J)_{a}^{~b}(V^KV^L)_{c}^{~d}(V^M)_{e}^{~f}(V^N)_{g}^{~h}$\\
        \hline
        \multirow{4}{9em}{Two $\bf{27}$, Seven $\bf{8}$}&$W_{fb}^{~~ak}W_{kd}^{~~ce}(V^IV^JV^K)_{a}^{~b}(V^LV^M)_{c}^{~d}(V^NV^O)_{e}^{~f}$\\
        &$W_{fb}^{~~ak}W_{kd}^{~~ce}(V^IV^J)_{a}^{~b}(V^KV^LV^M)_{c}^{~d}(V^NV^O)_{e}^{~f}$\\
         &$W_{fb}^{~~ag}W_{hd}^{~~ce}(V^IV^JV^K)_{a}^{~b}(V^LV^M)_{c}^{~d}(V^N)_{e}^{~f}(V^O)_{g}^{~h}$\\
         &$W_{fb}^{~~ag}W_{hd}^{~~ce}(V^IV^J)_{a}^{~b}(V^KV^L)_{c}^{~d}(V^MV^N)_{e}^{~f}(V^O)_{g}^{~h}$\\
        \hline
        \multirow{1}{9em}{Two $\bf{27}$, Eight $\bf{8}$}&$W_{fh}^{~~ac}W_{bd}^{~~eg}(V^IV^J)_{a}^{~b}(V^KV^L)_{c}^{~d}(V^MV^N)_{e}^{~f}(V^OV^P)_{g}^{~h}$\\
        \hline
\end{tabular}
\end{center}
\begin{center}
\renewcommand{\arraystretch}{1.25}
    \begin{tabular}{|c|c|}
        \hline
         &3HDM Invariants \\
        \hline
        \multirow{2}{9em}{Three $\bf{27}$, Zero $\bf{8}$} &$W_{ij}^{~~kl}W_{kl}^{~~ab}W_{ab}^{~~ij}$\\
        &$W_{ij}^{~~kl}W_{ka}^{~~ib}W_{lb}^{~~ja}$\\
        \hline
        \multirow{2}{9em}{Three $\bf{27}$, One $\bf{8}$} &$W_{ib}^{~~ak}W_{kj}^{~~lm}W_{ml}^{~~ij}(V^I)_{a}^{~b}$\\
        &$W_{bi}^{~~kl}W_{jm}^{~~ai}W_{kl}^{~~jm}(V^I)_{a}^{~b}$\\
        \hline
        \multirow{8}{9em}{Three $\bf{27}$, Two $\bf{8}$ }&$W_{bd}^{~~kl}W_{ij}^{~~ac}W_{kl}^{~~ij}(V^I)_{a}^{~b}(V^J)_{c}^{~d}$\\
        &$W_{ib}^{~~ak}W_{jd}^{~~cl}W_{kl}^{~~ij}(V^I)_{a}^{~b}(V^J)_{c}^{~d}$\\
        &$W_{ib}^{~~ck}W_{jd}^{~~al}W_{kl}^{~~ij}(V^I)_{a}^{~b}(V^J)_{c}^{~d}$\\
        &$W_{ib}^{~~kl}W_{jd}^{~~ac}W_{kl}^{~~ij}(V^I)_{a}^{~b}(V^J)_{c}^{~d}$\\
        &$W_{ij}^{~~ka}W_{bd}^{~~cl}W_{kl}^{~~ij}(V^I)_{a}^{~b}(V^J)_{c}^{~d}$\\
        &$W_{bd}^{~~kl}W_{ki}^{~~aj}W_{lj}^{~~ci}(V^I)_{a}^{~b}(V^J)_{c}^{~d}$\\
        &$W_{ib}^{~~ak}W_{kd}^{~~lj}W_{lj}^{~~ci}(V^I)_{a}^{~b}(V^J)_{c}^{~d}$\\
        &$W_{ib}^{~~ck}W_{kd}^{~~lj}W_{lj}^{~~ai}(V^I)_{a}^{~b}(V^J)_{c}^{~d}$\\
        \hline
        \multirow{14}{9em}{Three $\bf{27}$, Three $\bf{8}$} &$W_{bd}^{~~kl}W_{ij}^{~~ac}W_{kl}^{~~ij}(V^IV^J)_{a}^{~b}(V^K)_{c}^{~d}$\\
        &$W_{ib}^{~~kl}W_{jd}^{~~ac}W_{kl}^{~~ij}(V^I)_{a}^{~b}(V^JV^K)_{c}^{~d}$\\
        &$W_{bf}^{~~ac}W_{id}^{~~kl}W_{kl}^{~~ei}(V^I)_{a}^{~b}(V^J)_{c}^{~d}(V^K)_{e}^{~f}$\\
    &$W_{bd}^{~~le}W_{ij}^{~~ac}W_{fl}^{~~ij}(V^I)_{a}^{~b}(V^J)_{c}^{~d}(V^K)_{e}^{~f}$\\
    &$W_{bd}^{~~la}W_{ij}^{~~ec}W_{fl}^{~~ij}(V^I)_{a}^{~b}(V^J)_{c}^{~d}(V^K)_{e}^{~f}$\\
    &$W_{bd}^{~~al}W_{if}^{~~ck}W_{kl}^{~~ei}(V^I)_{a}^{~b}(V^J)_{c}^{~d}(V^K)_{e}^{~f}$\\
    &$W_{bd}^{~~al}W_{if}^{~~ek}W_{kl}^{~~ci}(V^I)_{a}^{~b}(V^J)_{c}^{~d}(V^K)_{e}^{~f}$\\
    &$W_{bd}^{~~el}W_{if}^{~~ak}W_{kl}^{~~ci}(V^I)_{a}^{~b}(V^J)_{c}^{~d}(V^K)_{e}^{~f}$\\
    &$W_{ib}^{~~ac}W_{df}^{~~kl}W_{kl}^{~~ei}(V^I)_{a}^{~b}(V^J)_{c}^{~d}(V^K)_{e}^{~f}$\\
    &$W_{if}^{~~ac}W_{db}^{~~kl}W_{kl}^{~~ei}(V^I)_{a}^{~b}(V^J)_{c}^{~d}(V^K)_{e}^{~f}$\\
    &$W_{ib}^{~~ac}W_{jf}^{~~ek}W_{dk}^{~~ij}(V^I)_{a}^{~b}(V^J)_{c}^{~d}(V^K)_{e}^{~f}$\\
    &$W_{if}^{~~ac}W_{jb}^{~~ek}W_{dk}^{~~ij}(V^I)_{a}^{~b}(V^J)_{c}^{~d}(V^K)_{e}^{~f}$\\
    &$W_{bd}^{~~kl}W_{fk}^{~~ei}W_{li}^{~~ac}(V^I)_{a}^{~b}(V^J)_{c}^{~d}(V^K)_{e}^{~f}$\\
    &$W_{bd}^{~~kl}W_{fk}^{~~ai}W_{li}^{~~ec}(V^I)_{a}^{~b}(V^J)_{c}^{~d}(V^K)_{e}^{~f}$\\
        \hline
\end{tabular}
\end{center}
\begin{center}
\renewcommand{\arraystretch}{1.25}
    \begin{tabular}{|c|c|}
        \hline
         &3HDM Invariants \\
        \hline
        \multirow{18}{8em}{Three $\bf{27}$, Four $\bf{8}$} &$W_{ib}^{~~ak}W_{jd}^{~~cl}W_{kl}^{~~ij}(V^1V^1)_{a}^{~b}(V^1V^1)_{c}^{~d}$\\
    &$W_{ib}^{~~kl}W_{jd}^{~~ac}W_{kl}^{~~ij}(V^IV^J)_{a}^{~b}(V^KV^L)_{c}^{~d}$\\
    &$W_{fb}^{~~ac}W_{id}^{~~kl}W_{kl}^{~~ei}(V^I)_{a}^{~b}(V^JV^K)_{c}^{~d}(V^L)_{e}^{~f}$\\
    &$W_{nd}^{~~le}W_{ij}^{~~ac}W_{fl}^{~~ij}(V^IV^J)_{a}^{~b}(V^K)_{c}^{~d}(V^L)_{e}^{~f}$\\
    &$W_{bd}^{~~le}W_{ij}^{~~ac}W_{fl}^{~~ij}(V^I)_{a}^{~b}(V^J)_{c}^{~d}(V^KV^L)_{e}^{~f}$\\
    &$W_{bd}^{~~la}W_{ij}^{~~ec}W_{fl}^{~~ij}(V^IV^J)_{a}^{~b}(V^K)_{c}^{~d}(V^L)_{e}^{~f}$\\
    &$W_{bf}^{~~ac}W_{dh}^{~~kl}W_{kl}^{~~eg}(V^I)_{a}^{~b}(V^J)_{c}^{~d}(V^K)_{e}^{~f}(V^L)_{g}^{~h}$\\
    &$W_{hf}^{~~ac}W_{db}^{~~kl}W_{kl}^{~~eg}(V^I)_{a}^{~b}(V^J)_{c}^{~d}(V^K)_{e}^{~f}(V^L)_{g}^{~h}$\\
    &$W_{bd}^{~~ae}W_{if}^{~~ck}W_{kh}^{~~gi}(V^I)_{a}^{~b}(V^J)_{c}^{~d}(V^K)_{e}^{~f}(V^L)_{g}^{~h}$\\
    &$W_{bd}^{~~ge}W_{if}^{~~ak}W_{kh}^{~~ci}(V^I)_{a}^{~b}(V^J)_{c}^{~d}(V^K)_{e}^{~f}(V^L)_{g}^{~h}$\\
    &$W_{bd}^{~~ai}W_{fj}^{~~eg}W_{hi}^{~~cj}(V^I)_{a}^{~b}(V^J)_{c}^{~d}(V^K)_{e}^{~f}(V^L)_{g}^{~h}$\\
    &$W_{bd}^{~~ai}W_{fj}^{~~cg}W_{hi}^{~~ej}(V^I)_{a}^{~b}(V^J)_{c}^{~d}(V^K)_{e}^{~f}(V^L)_{g}^{~h}$\\
    &$W_{bd}^{~~ei}W_{fj}^{~~ac}W_{hi}^{~~gj}(V^I)_{a}^{~b}(V^J)_{c}^{~d}(V^K)_{e}^{~f}(V^L)_{g}^{~h}$\\
    &$W_{bd}^{~~ei}W_{fj}^{~~ag}W_{hi}^{~~cj}(V^I)_{a}^{~b}(V^J)_{c}^{~d}(V^K)_{e}^{~f}(V^L)_{g}^{~h}$\\
    &$W_{bd}^{~~ai}W_{fh}^{~~cj}W_{ij}^{~~eg}(V^I)_{a}^{~b}(V^J)_{c}^{~d}(V^K)_{e}^{~f}(V^L)_{g}^{~h}$\\
    &$W_{bd}^{~~ai}W_{fh}^{~~ej}W_{ij}^{~~cg}(V^I)_{a}^{~b}(V^J)_{c}^{~d}(V^K)_{e}^{~f}(V^L)_{g}^{~h}$\\
    &$W_{bd}^{~~ei}W_{fh}^{~~aj}W_{ij}^{~~cg}(V^I)_{a}^{~b}(V^J)_{c}^{~d}(V^K)_{e}^{~f}(V^L)_{g}^{~h}$\\
    &$W_{ib}^{~~ac}W_{jf}^{~~eg}W_{dh}^{~~ij}(V^I)_{a}^{~b}(V^J)_{c}^{~d}(V^K)_{e}^{~f}(V^L)_{g}^{~h}$\\
        \hline
    \multirow{16}{8em}{Three $\bf{27}$, Five $\bf{8}$} &$W_{bd}^{~~le}W_{ij}^{~~ac}W_{fl}^{~~ij}(V^I)_{a}^{~b}(V^JV^K)_{c}^{~d}(V^LV^M)_{e}^{~f}$\\
    &$W_{fb}^{~~ac}W_{id}^{~~kl}W_{kl}^{~~ei}(V^IV^J)_{a}^{~b}(V^K)_{c}^{~d}(V^LV^M)_{e}^{~f}$\\
    &$W_{bf}^{~~ac}W_{dh}^{~~kl}W_{kl}^{~~eg}(V^I)_{a}^{~b}(V^JV^K)_{c}^{~d}(V^L)_{e}^{~f}(V^M)_{g}^{~h}$\\
    &$W_{hf}^{~~ac}W_{db}^{~~kl}W_{kl}^{~~eg}(V^I)_{a}^{~b}(V^JV^K)_{c}^{~d}(V^L)_{e}^{~f}(V^M)_{g}^{~h}$\\
    &$W_{bd}^{~~ae}W_{if}^{~~ck}W_{kl}^{~~gi}(V^I)_{a}^{~b}(V^JV^K)_{c}^{~d}(V^L)_{e}^{~f}(V^M)_{g}^{~h}$\\
    &$W_{bd}^{~~ce}W_{if}^{~~gk}W_{kh}^{~~ai}(V^I)_{a}^{~b}(V^JV^K)_{c}^{~d}(V^L)_{e}^{~f}(V^M)_{g}^{~h}$\\
    &$W_{bd}^{~~ge}W_{if}^{~~ak}W_{kh}^{~~ci}(V^I)_{a}^{~b}(V^JV^K)_{c}^{~d}(V^L)_{e}^{~f}(V^M)_{g}^{~h}$\\
    &$W_{bd}^{~~ai}W_{fj}^{~~ce}W_{hi}^{~~gj}(V^I)_{a}^{~b}(V^JV^K)_{c}^{~d}(V^L)_{e}^{~f}(V^M)_{g}^{~h}$\\
    &$W_{bd}^{~~ai}W_{fj}^{~~eg}W_{hi}^{~~cj}(V^I)_{a}^{~b}(V^JV^K)_{c}^{~d}(V^L)_{e}^{~f}(V^M)_{g}^{~h}$\\
    &$W_{bd}^{~~ai}W_{fh}^{~~cj}W_{ij}^{~~eg}(V^I)_{a}^{~b}(V^JV^K)_{c}^{~d}(V^L)_{e}^{~f}(V^M)_{g}^{~h}$\\
    &$W_{bd}^{~~ai}W_{fh}^{~~ej}W_{ij}^{~~cg}(V^I)_{a}^{~b}(V^JV^K)_{c}^{~d}(V^L)_{e}^{~f}(V^M)_{g}^{~h}$\\
    &$W_{bd}^{~~ae}W_{fh}^{~~ck}W_{jk}^{~~gi}(V^I)_{a}^{~b}(V^J)_{c}^{~d}(V^K)_{e}^{~f}(V^L)_{g}^{~h}(V^M)_{i}^{~j}$\\
    &$W_{bd}^{~~ae}W_{fh}^{~~ik}W_{jk}^{~~cg}(V^I)_{a}^{~b}(V^J)_{c}^{~d}(V^K)_{e}^{~f}(V^L)_{g}^{~h}(V^M)_{i}^{~j}$\\
    &$W_{bd}^{~~eg}W_{fh}^{~~ak}W_{jk}^{~~ci}(V^I)_{a}^{~b}(V^J)_{c}^{~d}(V^K)_{e}^{~f}(V^L)_{g}^{~h}(V^M)_{i}^{~j}$\\
    &$W_{bd}^{~~eg}W_{fh}^{~~ik}W_{jk}^{~~ac}(V^I)_{a}^{~b}(V^J)_{c}^{~d}(V^K)_{e}^{~f}(V^L)_{g}^{~h}(V^M)_{i}^{~j}$\\
    &$W_{bd}^{~~ai}W_{fh}^{~~ck}W_{jk}^{~~eg}(V^I)_{a}^{~b}(V^J)_{c}^{~d}(V^K)_{e}^{~f}(V^L)_{g}^{~h}(V^M)_{i}^{~j}$\\
    \hline
    \end{tabular}
\end{center}
\begin{center}
\renewcommand{\arraystretch}{1.25}
    \begin{tabular}{|c|c|}
        \hline
         &3HDM Invariants \\
        \hline
        \multirow{16}{9em}{Three $\bf{27}$, Six $\bf{8}$} &$W_{bd}^{~~le}W_{ij}^{~~ac}W_{fl}^{~~ij}(V^IV^J)_{a}^{~b}(V^KV^L)_{c}^{~d}(V^MV^L)_{e}^{~f}$\\
    &$W_{bd}^{~~al}W_{if}^{~~ck}W_{kl}^{~~ei}(V^IV^J)_{a}^{~b}(V^KV^L)_{c}^{~d}(V^MV^L)_{e}^{~f}$\\
    &$W_{ib}^{~~ac}W_{df}^{~~kl}W_{kl}^{~~ei}(V^IV^J)_{a}^{~b}(V^KV^L)_{c}^{~d}(V^MV^L)_{e}^{~f}$\\
    &$W_{ib}^{~~ac}W_{jf}^{~~ek}W_{dk}^{~~ij}(V^IV^J)_{a}^{~b}(V^KV^L)_{c}^{~d}(V^MV^L)_{e}^{~f}$\\
    &$W_{bf}^{~~ac}W_{dh}^{~~kl}W_{kl}^{~~eg}(V^IV^J)_{a}^{~b}(V^KV^L)_{c}^{~d}(V^M)_{e}^{~f}(V^N)_{g}^{~h}$\\
    &$W_{hf}^{~~ac}W_{db}^{~~kl}W_{kl}^{~~eg}(V^IV^J)_{a}^{~b}(V^KV^L)_{c}^{~d}(V^M)_{e}^{~f}(V^N)_{g}^{~h}$\\
    &$W_{bd}^{~~ae}W_{if}^{~~ck}W_{kh}^{~~gi}(V^IV^J)_{a}^{~b}(V^KV^L)_{c}^{~d}(V^M)_{e}^{~f}(V^N)_{g}^{~h}$\\
    &$W_{bd}^{~~ae}W_{fh}^{~~ck}W_{jk}^{~~gi}(V^IV^J)_{a}^{~b}(V^K)_{c}^{~d}(V^L)_{e}^{~f}(V^M)_{g}^{~h}(V^N)_{i}^{~j}$\\
    &$W_{bd}^{~~ae}W_{fh}^{~~gk}W_{jk}^{~~ci}(V^IV^J)_{a}^{~b}(V^K)_{c}^{~d}(V^L)_{e}^{~f}(V^M)_{g}^{~h}(V^N)_{i}^{~j}$\\
    &$W_{bd}^{~~eg}W_{fh}^{~~mk}W_{jk}^{~~ci}(V^IV^J)_{a}^{~b}(V^K)_{c}^{~d}(V^L)_{e}^{~f}(V^M)_{g}^{~h}(V^N)_{i}^{~j}$\\
    &$W_{bd}^{~~eg}W_{fh}^{~~ik}W_{jk}^{~~ac}(V^IV^J)_{a}^{~b}(V^K)_{c}^{~d}(V^L)_{e}^{~f}(V^M)_{g}^{~h}(V^N)_{i}^{~j}$\\
    &$W_{bd}^{~~ai}W_{fh}^{~~ck}W_{jk}^{~~eg}(V^IV^J)_{a}^{~b}(V^K)_{c}^{~d}(V^L)_{e}^{~f}(V^M)_{g}^{~h}(V^N)_{i}^{~j}$\\
    &$W_{bd}^{~~ei}W_{fh}^{~~ak}W_{jk}^{~~cg}(V^IV^J)_{a}^{~b}(V^K)_{c}^{~d}(V^L)_{e}^{~f}(V^M)_{g}^{~h}(V^N)_{i}^{~j}$\\
    &$W_{bd}^{~~ae}W_{fh}^{~~gk}W_{jk}^{~~ci}(V^I)_{a}^{~b}(V^J)_{c}^{~d}(V^KV^L)_{e}^{~f}(V^M)_{g}^{~h}(V^N)_{i}^{~j}$\\
    &$W_{bd}^{~~ik}W_{fh}^{~~ac}W_{jl}^{~~eg}(V^I)_{a}^{~b}(V^J)_{c}^{~d}(V^K)_{e}^{~f}(V^L)_{g}^{~h}(V^M)_{i}^{~j}(V^N)_{k}^{~l}$\\
    &$W_{db}^{~~ie}W_{fh}^{~~ak}W_{jl}^{~~cg}(V^I)_{a}^{~b}(V^J)_{c}^{~d}(V^K)_{e}^{~f}(V^L)_{g}^{~h}(V^M)_{i}^{~j}(V^N)_{k}^{~l}$\\
        \hline
    \multirow{6}{9em}{Three $\bf{27}$, Seven $\bf{8}$}  &$W_{hf}^{~~ac}W_{db}^{~~kl}W_{kl}^{~~eg}(V^I)_{a}^{~b}(V^JV^K)_{c}^{~d}(V^LV^M)_{e}^{~f}(V^NV^O)_{g}^{~h}$\\
    &$W_{bd}^{~~ce}W_{if}^{~~gk}W_{kh}^{~~ai}(V^I)_{a}^{~b}(V^JV^K)_{c}^{~d}(V^LV^M)_{e}^{~f}(V^NV^O)_{g}^{~h}$\\
    &$W_{bd}^{~~ai}W_{fj}^{~~ce}W_{hi}^{~~gj}(V^I)_{a}^{~b}(V^JV^K)_{c}^{~d}(V^LV^M)_{e}^{~f}(V^NV^O)_{g}^{~h}$\\
    &$W_{bd}^{~~ai}W_{fh}^{~~cj}W_{ij}^{~~eg}(V^I)_{a}^{~b}(V^JV^K)_{c}^{~d}(V^LV^M)_{e}^{~f}(V^NV^O)_{g}^{~h}$\\
    &$W_{bd}^{~~ie}W_{fh}^{~~ak}W_{jl}^{~~cg}(V^I)_{a}^{~b}(V^JV^K)_{c}^{~d}(V^L)_{e}^{~f}(V^M)_{g}^{~h}(V^N)_{i}^{~j}(V^O)_{k}^{~l}$\\
    &$W_{bd}^{~~ie}W_{fh}^{~~ak}W_{jl}^{~~cg}(V^I)_{a}^{~b}(V^J)_{c}^{~d}(V^KV^L)_{e}^{~f}(V^M)_{g}^{~h}(V^N)_{i}^{~j}(V^O)_{k}^{~l}$\\
    \hline
    \multirow{2}{9em}{Three $\bf{27}$, Eight $\bf{8}$} &$W_{bd}^{~~ai}W_{fh}^{~~cj}W_{ij}^{~~eg}(V^IV^J)_{a}^{~b}(V^KV^L)_{c}^{~d}(V^MV^N)_{e}^{~f}(V^OV^P)_{g}^{~h}$\\
    &$W_{db}^{~~ie}W_{fh}^{~~ak}W_{jl}^{~~cg}(V^IV^J)_{a}^{~b}(V^KV^L)_{c}^{~d}(V^M)_{e}^{~f}(V^N)_{g}^{~h}(V^O)_{i}^{~j}(V^P)_{k}^{~l}$\\
    \hline
    \multirow{1}{9em}{Three $\bf{27}$, Nine $\bf{8}$ }&$W_{bd}^{~~ai}W_{fh}^{~~ck}W_{jk}^{~~eg}(V^IV^J)_{a}^{~b}(V^KV^L)_{c}^{~d}(V^MV^N)_{e}^{~f}(V^OV^P)_{g}^{~h}(V^Q)_{i}^{~j}$\\
    \hline
    \end{tabular}
\end{center}

\section{CP Properties of the Invariants}\label{CPinvs}
As a first application of our results, we investigate the CP properties of the invariants appearing in the 3HDM. In particular, we classify the invariants into CP-even and CP-odd combinations.
{This classification is purely algebraic: it refers to the transformation of the invariant polynomials under the CP transformation, rather than by itself implying the presence of physical CP violation.}
We first carry out this analysis in the $U(1)_1\times U(1)_2$ basis obtained after gauge fixing, where the CP transformation properties are particularly transparent. We then use these results to determine the corresponding CP-even and CP-odd invariants of the full $SU(3)$ theory. We carry out this analysis for the invariants identified up to four fields. A complete classification of CP-even and CP-odd invariants is left for future work, together with a determination of the full invariant ring.

\subsection{CP Building Blocks}
Following~\cite{Trautner:2018ipq}, we begin with the most general CP transformation acting on the Higgs fields. Suppressing the $SU(2)_L$ indices, the transformation takes the form
\begin{align}
    \Phi_i &\xrightarrow{\mathrm{CP}} (\Phi^*)^j [U^T]_j^{~i}\\
    (\Phi^*)^i &\xrightarrow{\mathrm{CP}}  [U^*]_i^{~j}\Phi_j
\end{align}
where $U$ is a unitary matrix acting in flavor space.

To determine the transformation properties of the mass coupling, consider
\begin{align}
    (\Phi^*)^j(\mu^2)^{~i}_{j}\Phi_i\xrightarrow{\mathrm{CP}} &[U^*]^{~k}_j\Phi_k\,(\mu^2)^{~i}_{j} (\Phi^*)^l [U^T]_l^{~i}\\
    &=(\Phi^*)^l [U^T]_l^{~i}(\mu^2)^{~i}_{j} [U^*]^{~k}_j(\Phi)_k\\
    &=(\Phi^*)^j ([U^T](\mu^2)^T [U^*])_j^{~i}\Phi_i\,.
\end{align}

Hence,
\begin{align}
    (\mu^2) \xrightarrow{\mathrm{CP}} (U^T (\mu^2)^T U^*)\,.
\end{align}

Similarly, for the quartic coupling, one obtains
\begin{align}
    \lambda_{ij}^{~kl}\xrightarrow{\mathrm{CP}} [U^T]_i^n [U^T]_j^{n'}(\lambda^T)_{nn'}^{~mm'}[U^*]_m^k [U^*]_{m'}^l\,.
\end{align}

Restricting ourselves to CP transformations of order two, one may always
choose a basis in flavor space in which
\begin{align}
    U \propto \mathbf{1}.
\end{align}
Working in such a basis, the transformation properties of the coupling
tensors simplify to
\begin{align}
    (\mu^2)_i^{~j}&\xrightarrow{\mathrm{CP}} (\mu^2)_j^{~i}=((\mu^2)^*)_i^{~j}\\
    \lambda_{ij}^{~kl}&\xrightarrow{\mathrm{CP}} \lambda_{kl}^{~ij}=(\lambda^*)_{ij}^{~kl}
\end{align}
where the final equalities follow from Hermiticity.

Using the definitions of the adjoint fields and the $\mathbf{27}$ in terms of $\mu^2$ and $\lambda$, it is straightforward to show that CP acts on the irreducible $SU(3)$ building blocks by exchanging fundamental and antifundamental flavor indices. The resulting transformation properties are
\begin{align}
(V^{I})_i^{~j} &\xrightarrow{\mathrm{CP}} (V^{I})_j^{~i}\,\\
W_{ij}^{~~kl} &\xrightarrow{\mathrm{CP}} W_{kl}^{~~ij}\,.
\end{align}

Consequently, determining the CP properties of an invariant reduces to studying how it transforms under the exchange of upper and lower flavor indices.

As a simple example, consider the invariant
\begin{align}
W_{ij}^{~~kl}V_k^{~i}V_l^{~j}.
\end{align}
Since all fundamental and antifundamental indices are contracted pairwise, this object is invariant under the exchange of upper and lower indices and is therefore CP even.

More generally, however, tensor structures need not possess this symmetry. For example,
\begin{align}
W_{ij}^{~~ka}W_{bd}^{~~cl}W_{kl}^{~~ij}
(V^I)_a^{~b}(V^J)_c^{~d}
\end{align}
is not manifestly invariant under the exchange of fundamental and antifundamental indices. Such structures are therefore not necessarily CP even by themselves and may instead be combined to form either CP-even or CP-odd invariants.

\subsection{CP on $U(1)_1\times U(1)_2$ Invariants}
It is often more convenient to determine the CP properties of invariants using the $U(1)_1\times U(1)_2$ building blocks introduced in Section~\ref{u1u1invariants}. Before doing so, we must verify that the gauge fixing used to obtain these building blocks does not itself break CP. Since
\begin{align}
    \langle V^1 \rangle &\xrightarrow{\mathrm{CP}} \langle V^1 \rangle ^T =\langle V^1 \rangle
\end{align}
the chosen vacuum configuration is CP invariant, as $\langle V^1\rangle$ is diagonal.

Since the gauge fixing preserves CP, the CP properties of the $U(1)_1\times U(1)_2$ invariants may be used to determine the CP properties of the corresponding 3HDM invariants. If a $U(1)_1\times U(1)_2$ field component carries charge $Q_{q_1,q_2}$, CP acts by exchanging all fundamental and antifundamental indices. Consequently,
\begin{align}
    Q_{q_1,q_2}&\xrightarrow{\mathrm{CP}} Q_{-q_1,-q_2}\,.
\end{align}

We first consider invariants quadratic in the field content. Such invariants must be of the form
\begin{align}
Q_{q_1,q_2}Q_{-q_1,-q_2},
\end{align}
in order to be neutral under $U(1)_1\times U(1)_2$.
In some cases the field components associated with the charges $Q_{q_1,q_2}$ and $Q_{-q_1,-q_2}$ are already related by CP. An example is
\begin{align}
(V^2)_2^{~1}(V^2)_1^{~2},
\end{align}
which is manifestly CP even. In general, however, the two field components need not be CP conjugates of one another. Examples include
\begin{align}
W_{12}^{~~11}W_{12}^{~~22},
\qquad
W_{12}^{~~11}(V^2)_1^{~2}.
\end{align}
In such cases, CP-even and CP-odd invariants are obtained by taking the symmetric and antisymmetric combinations, respectively.

We therefore define the quadratic CP-even and CP-odd invariants
\begin{align}
    \mathcal{I}^{I,J}_+&=Q^{(I}_{q_1,q_2}Q^{J)}_{-q_1,-q_2}\\
    \mathcal{I}^{I,J}_-&=Q^{[I}_{q_1,q_2}Q^{J]}_{-q_1,-q_2}
\end{align}
where $Q^I_{q_1,q_2}$ denotes any field component carrying charge $(q_1,q_2)$. We list out all of these invariants below.
\begin{center}
\renewcommand{\arraystretch}{1.25}
    \begin{tabular}{|c|c|c|}
        \hline
        $U(1)_1\times U(1)_2$ Invariant & $\mathcal{I}^{I,J}_+$  & $\mathcal{I}^{I,J}_-$\\
        \hline
        \multirow{6}{5em}{$Q_{-1, -1}Q_{1, 1}$}&$(V^{(I})_{1}^{~2}(V^{J)})_{2}^{~1}$&$(V^{[I})_{1}^{~2}(V^{J]})_{2}^{~1}$\\&$W_{12}^{~~11}(V^I)_{1}^{~2}+W_{11}^{~~12}(V^I)_{2}^{~1}$&$W_{12}^{~~11}(V^I)_{1}^{~2}-W_{11}^{~~12}(V^I)_{2}^{~1}$\\&$W_{22}^{~~12}(V^I)_{1}^{~2}+W_{12}^{~~22}(V^I)_{2}^{~1}$&$W_{22}^{~~12}(V^I)_{1}^{~2}-W_{12}^{~~22}(V^I)_{2}^{~1}$\\&$W_{11}^{~~12}W_{22}^{~~12}+W_{12}^{~~22}W_{12}^{~~11}$&$W_{11}^{~~12}W_{22}^{~~12}-W_{12}^{~~22}W_{12}^{~~11}$\\&$W_{11}^{~~12}W_{12}^{~~11}$&--\\&$W_{12}^{~~22}W_{22}^{~~12}$&--\\\hline\multirow{6}{5em}{$Q_{1, -1}Q_{-1, 1}$}&$(V^{(I})_{1}^{~3}(V^{J)})_{3}^{~1}$&$(V^{[I})_{1}^{~3}(V^{J]})_{3}^{~1}$\\&$W_{13}^{~~11}(V^I)_{1}^{~3}+W_{11}^{~~13}(V^I)_{3}^{~1}$&$W_{13}^{~~11}(V^I)_{1}^{~3}-W_{11}^{~~13}(V^I)_{3}^{~1}$\\&$W_{23}^{~~12}(V^I)_{1}^{~3}+W_{12}^{~~23}(V^I)_{3}^{~1}$&$W_{23}^{~~12}(V^I)_{1}^{~3}-W_{12}^{~~23}(V^I)_{3}^{~1}$\\&$W_{11}^{~~13}W_{23}^{~~12}+W_{12}^{~~23}W_{13}^{~~11}$&$W_{11}^{~~13}W_{23}^{~~12}-W_{12}^{~~23}W_{13}^{~~11}$\\&$W_{11}^{~~13}W_{13}^{~~11}$&--\\&$W_{12}^{~~23}W_{23}^{~~12}$&--\\\hline\multirow{6}{5em}{$Q_{2, 0}Q_{-2, 0}$}&$(V^{(I})_{2}^{~3}(V^{J)})_{3}^{~2}$&$(V^{[I})_{2}^{~3}(V^{J]})_{3}^{~2}$\\&$W_{13}^{~~12}(V^I)_{2}^{~3}+W_{12}^{~~13}(V^I)_{3}^{~2}$&$W_{13}^{~~12}(V^I)_{2}^{~3}-W_{12}^{~~13}(V^I)_{3}^{~2}$\\&$W_{23}^{~~22}(V^I)_{2}^{~3}+W_{22}^{~~23}(V^I)_{3}^{~2}$&$W_{23}^{~~22}(V^I)_{2}^{~3}-W_{22}^{~~23}(V^I)_{3}^{~2}$\\&$W_{12}^{~~13}W_{23}^{~~22}+W_{22}^{~~23}W_{13}^{~~12}$&$W_{12}^{~~13}W_{23}^{~~22}-W_{22}^{~~23}W_{13}^{~~12}$\\&$W_{12}^{~~13}W_{13}^{~~12}$&--\\&$W_{22}^{~~23}W_{23}^{~~22}$&--\\\hline
        \multirow{1}{5em}{$Q_{-3, -1}Q_{3, 1}$}&$W_{13}^{~~22}W_{22}^{~~13}$&--\\\hline\multirow{1}{5em}{$Q_{-2, -2}Q_{2, 2}$}&$W_{11}^{~~22}W_{22}^{~~11}$&--\\\hline\multirow{1}{5em}{$Q_{0, -2}Q_{0, 2}$}&$W_{11}^{~~23}W_{23}^{~~11}$&--\\\hline\multirow{1}{5em}{$Q_{2, -2}Q_{-2, 2}$}&$W_{11}^{~~33}W_{33}^{~~11}$&--\\\hline\multirow{1}{5em}{$Q_{3, -1}Q_{-3, 1}$}&$W_{12}^{~~33}W_{33}^{~~12}$&--\\\hline\multirow{1}{5em}{$Q_{4, 0}Q_{-4, 0}$}&$W_{22}^{~~33}W_{33}^{~~22}$&--\\\hline
\end{tabular}
\end{center}

The case of $U(1)_1\times U(1)_2$ invariants involving three or more fields is slightly different. Consider an invariant constructed from field components with charges
\begin{align}
Q_{q_1^1,q_2^1}\cdots Q_{q_1^n,q_2^n}\,,
\end{align}
where
\begin{align}
\sum_{i=1}^{n}(q_1^i,q_2^i)=(0,0)\,.
\end{align}

For an independent invariant, no charge $Q_{q_1,q_2}$ can appear simultaneously with its conjugate charge $Q_{-q_1,-q_2}$. Indeed, if both charges were present, the invariant would factorize as
\begin{align}
Q_{q_1,q_2}Q_{-q_1,-q_2}
\end{align}
times a lower-order $U(1)_1\times U(1)_2$ invariant and would therefore not be independent.

Consequently, under CP the invariant is mapped to a distinct invariant obtained by reversing the sign of every charge,
\begin{align}
Q_{q_1^1,q_2^1}\cdots Q_{q_1^n,q_2^n}
\xrightarrow{\mathrm{CP}}
Q_{-q_1^1,-q_2^1}\cdots Q_{-q_1^n,-q_2^n}.
\end{align}
CP-even and CP-odd combinations are therefore obtained by taking the symmetric and antisymmetric linear combinations
\begin{align}
    \mathcal{I}_+&=Q_{q^1_1,q^1_2} \dots Q_{q^n_1,q^n_2}+ Q_{-q^1_1,-q^1_2}\cdot \dots Q_{-q^n_1,-q^n_2}\\
    \mathcal{I}_-&=Q_{q^1_1,q^1_2} \dots Q_{q^n_1,q^n_2}- Q_{-q^1_1,-q^1_2} \dots Q_{-q^n_1,-q^n_2}\,.
\end{align}
{One can build all CP-even and CP-odd $U(1)_1\times U(1)_2$ invariants by utilizing the tables listed in the Appendix. }

Having classified the $U(1)_1\times U(1)_2$ invariants, the CP properties of the corresponding 3HDM invariants may then be determined either by matching them to the $U(1)_1\times U(1)_2$ basis, as described in Section~\ref{3hdmmatching}, or directly from the transformation properties
\begin{align}
(V^I)_i^{~j} &\xrightarrow{CP} (V^I)_j^{~i},
\\
W_{ij}^{~~kl} &\xrightarrow{CP} W_{kl}^{~~ij}
\end{align}
under CP.

We carried out this procedure for all independent invariants we discovered involving up to four fields. In the special case of invariants involving no $\bf{27}$ fields, i.e. only $\bf{8}$ fields, we applied the analysis to all the invariants. The resulting CP-even and CP-odd invariants are summarized below. For invariants involving four or more fields, CP-even and CP-odd combinations may similarly be constructed by symmetrizing and antisymmetrizing invariants related by CP. A complete classification of all CP-even and CP-odd 3HDM invariants is left for future work.
\begin{center}
\renewcommand{\arraystretch}{1.35}
\begin{tabular}{|c|c|c|}
\hline
Field content & CP-even invariants $\mathcal I_+$ & CP-odd invariants $\mathcal I_-$ \\
\hline
\multirow{1}{9em}{Zero $\mathbf{27}$, Two  $\mathbf{8}$}
&
$\mathrm{Tr}(V^I V^J)$
&
-- \\
\hline
\multirow{1}{9em}{Zero $\mathbf{27}$, Three  $\mathbf{8}$}
&
$\mathrm{Tr}(V^{(I}V^J V^{K)})$
&
$\mathrm{Tr}(V^{[I}V^J V^{K]})$ \\
\hline
\multirow{1}{9em}{Zero $\mathbf{27}$, Four  $\mathbf{8}$}
&
$\mathrm{Tr}(V^{[I}V^{J]} V^{[K}V^{L]})$
&
$\mathrm{Tr}(V^{[I}V^J V^{K]}V^L)$ \\
\hline
\multirow{1}{9em}{Zero $\mathbf{27}$, Five $\mathbf{8}$}
&
$\mathrm{Tr}([V^IV^JV^K][V^LV^M])$
&
$\mathrm{Tr}([V^IV^JV^K]V^{(L}V^{M)})$ \\
\hline
\multirow{1}{9em}{Zero $\mathbf{27}$, Six $\mathbf{8}$}
&
--
&
$\mathrm{Tr}([V^IV^J][V^KV^L][V^MV^N])$ \\
\hline
\multirow{1}{9em}{One $\mathbf{27}$, Two $\mathbf{8}$}
&
$W_{ij}^{~~kl}(V^{I})_k^{~i}(V^{J})_l^{~j}$
&
--
 \\
\hline
\multirow{1}{9em}{One $\mathbf{27}$, Three $\mathbf{8}$}
&
$W_{ij}^{~~kl}(V^{(I}V^{J)})_k^{~i}(V^{K})_l^{~j}$
&
$W_{ij}^{~~kl}(V^{[I}V^{J]})_k^{~i}(V^{K})_l^{~j}$
 \\
\hline
\multirow{1}{9em}{Two $\mathbf{27}$, Zero $\mathbf{8}$}
&
$W_{ij}^{~~kl}W_{kl}^{~~ij}$
&
--
 \\
\hline
\multirow{1}{9em}{Two $\mathbf{27}$, One $\mathbf{8}$}
&
$W_{ij}^{~~ka}W_{bk}^{~~ij}(V^I)_{a}^{~b}$
&
--
 \\
 \hline
\multirow{2}{9em}{Two $\mathbf{27}$, Two $\mathbf{8}$}
&
$W_{ij}^{~~ka}W_{bk}^{~~ij}(V^{(I}V^{J)})_{a}^{~b}$&$W_{ij}^{~~ka}W_{bk}^{~~ij}(V^{[I}V^{J]})_{a}^{~b}$\\
&
$W_{ib}^{~~ak}W_{kd}^{~~ci}(V^I)_{a}^{~b}(V^J)_{c}^{~d}$
&
--
 \\
\hline
\multirow{2}{9em}{Three $\mathbf{27}$, Zero $\mathbf{8}$}
&
$W_{ij}^{~~kl}W_{kl}^{~~ab}W_{ab}^{~~ij}$&--\\
&
$W_{ij}^{~~kl}W_{ka}^{~~ib}W_{lb}^{~~ja}$
&
--
 \\
\hline
\multirow{2}{9em}{Three $\mathbf{27}$, One $\mathbf{8}$}
&
$W_{ib}^{~~ak}W_{kj}^{~~lm}W_{ml}^{~~ij}(V^I)_{a}^{~b}$&--\\
&
$W_{bi}^{~~kl}W_{jm}^{~~ai}W_{kl}^{~~jm}(V^I)_{a}^{~b}$
&
--
 \\
\hline
\end{tabular}
\end{center}

\section{Conclusion}
In this work, we have studied aspects of the invariant structure of the 3HDM.

In the first place, we have computed the complete  Hilbert series of the 3HDM. Owing to the size 
of the 3HDM Hilbert series, standard computational techniques were insufficient, requiring the development of alternative methods to render the calculation tractable. We expect that these techniques will be useful more generally for the study of large  Hilbert series arising in other models with extended scalar sectors or nontrivial flavor symmetries.

Secondly,  we have presented  an approach to a  systematic classification of invariant operators built from the parameters of the 3HDM. Using these methods, we explicitly constructed a basis of independent invariant operators up to $\mathcal{O}(\lambda^3)$ in the quartic coupling. However, in the special case of invariants containing no $\mathbf{27}$ fields, the results are obtained to arbitrary order in the quartic coupling $\lambda$. While the complete classification of invariants at arbitrary order is straightforward, the operators identified here are expected to already capture the most phenomenologically relevant structures. In particular, they provide the necessary building blocks for the applications in 3HDM phenomenology and model building at low order in  $\lambda$.

{Finally, we applied our basis of invariant operators to begin the construction of CP-even and CP-odd invariants. We explicitly constructed CP-even and CP-odd combinations for invariants involving up to four fields and described how this procedure naturally extends to invariants of higher degree.
}

Several directions for future work naturally follow from our analysis. Firstly, one important extension would be the systematic construction of higher-order invariant operators beyond $\mathcal{O}(\lambda^3)$, which would further clarify the structure of the invariant ring of the 3HDM.  We estimate there are roughly over 2000 more independent $U(1)_1\times U(1)_2$ invariants to arbitrary order in $\lambda$. 

{An important remaining problem is the determination of the syzygies among the generators of the invariant ring. While the Hilbert series determines the counting of independent invariants and our explicit construction provides a basis through $\mathcal O(\lambda^3)$, the syzygies are required for a complete description of the invariant ring. Determining these relations would therefore be a necessary step toward a full characterization of the 3HDM parameter space in terms of basis-invariant quantities.
}

Another interesting direction would be to search for a more transparent or compact representation of the invariant operators identified here, which in their current form are somewhat cumbersome. 
 It would be useful to discover a more efficient method for representing these invariants.

The invariant operators identified in this work may also serve as useful inputs for phenomenological studies of CP violation, vacuum structure, and dark matter candidates in the 3HDM.
These invariants provide a basis-independent way to characterize the scalar sector of the 3HDM and can be used to identify physically distinct regions of parameter space. In particular, they may be employed to construct CP-odd invariants to search for explicit or spontaneous CP violation, as well as to study the symmetry structure of the scalar potential in a basis-independent manner. {We have initiated this program in the present work; however, it would be interesting to determine whether a general criterion exists for identifying a priori whether a given invariant operator is CP-even or CP-odd.} Such tools are particularly valuable when exploring the large parameter space of multi-Higgs models. We hope to report on these in future work.

Finally, 
it would also be interesting to apply the techniques developed here to other multi-Higgs models, including general $N$-Higgs-doublet models with $N>3$, where the complexity of the invariant structure grows rapidly. 

\section{Acknowledgments}
This work was supported in part by NSF PHY-2210283.

\section{Appendix}
Here we list all $U(1)_1\times U(1)_2$ invariants up to $\mathcal{O}(\lambda^3)$. The left column shows the charges of the fields, while the right column lists the corresponding invariant combinations of fields.
 Note that the $\lambda^3$ invariants consist of products of three fields. From the table of $U(1)_1\times U(1)_2$ charges, there is a symmetry between $Q_{q_1,q_2}$ and $Q_{-q_1,-q_2}$, which corresponds to replacing the fundamental indices of the fields with their anti-fundamental indices, and vice versa. 
\begin{center}
\renewcommand{\arraystretch}{1.25}
    \begin{tabular}{|c|c|}
        \hline
        Charges & $U(1)_1\times U(1)_2$ Invariants \\
        \hline
        \multirow{5}{5em}{ $Q_{0,0}$} & $(V^I)_{1}^{~1}$\\& $(V^I)_{2}^{~2}$\\& $W_{11}^{~~11}$\\ &$W_{12}^{~~12}$\\&$W_{22}^{~~22}$\\
        \hline
        \multirow{9}{5em}{$Q_{-1, -1}Q_{1, 1}$}&$(V^I)_{1}^{~2}(V^J)_{2}^{~1}$\\&$W_{12}^{~~11}(V^I)_{1}^{~2}$\\&$W_{22}^{~~12}(V^I)_{1}^{~2}$\\&$W_{11}^{~~12}(V^I)_{2}^{~1}$\\&$W_{12}^{~~22}(V^I)_{2}^{~1}$\\&$W_{11}^{~~12}W_{12}^{~~11}$\\&$W_{11}^{~~12}W_{22}^{~~12}$\\&$W_{12}^{~~22}W_{12}^{~~11}$\\&$W_{12}^{~~22}W_{22}^{~~12}$\\\hline\multirow{9}{5em}{$Q_{1, -1}Q_{-1, 1}$}&$(V^I)_{1}^{~3}(V^J)_{3}^{~1}$\\&$W_{13}^{~~11}(V^I)_{1}^{~3}$\\&$W_{23}^{~~12}(V^I)_{1}^{~3}$\\&$W_{11}^{~~13}(V^I)_{3}^{~1}$\\&$W_{12}^{~~23}(V^I)_{3}^{~1}$\\&$W_{11}^{~~13}W_{13}^{~~11}$\\&$W_{11}^{~~13}W_{23}^{~~12}$\\&$W_{12}^{~~23}W_{13}^{~~11}$\\&$W_{12}^{~~23}W_{23}^{~~12}$\\\hline\multirow{9}{5em}{$Q_{2, 0}Q_{-2, 0}$}&$(V^I)_{2}^{~3}(V^J)_{3}^{~2}$\\&$W_{13}^{~~12}(V^I)_{2}^{~3}$\\&$W_{23}^{~~22}(V^I)_{2}^{~3}$\\&$W_{12}^{~~13}(V^I)_{3}^{~2}$\\&$W_{22}^{~~23}(V^I)_{3}^{~2}$\\&$W_{12}^{~~13}W_{13}^{~~12}$\\&$W_{12}^{~~13}W_{23}^{~~22}$\\&$W_{22}^{~~23}W_{13}^{~~12}$\\&$W_{22}^{~~23}W_{23}^{~~22}$\\\hline
\end{tabular}
\end{center}
\begin{center}
\renewcommand{\arraystretch}{1.25}
    \begin{tabular}{|c|c|}
    \hline
        Charges & $U(1)_1\times U(1)_2$ Invariants \\
        \hline
        \multirow{1}{5em}{$Q_{-3, -1}Q_{3, 1}$}&$W_{13}^{~~22}W_{22}^{~~13}$\\\hline\multirow{1}{5em}{$Q_{-2, -2}Q_{2, 2}$}&$W_{11}^{~~22}W_{22}^{~~11}$\\\hline\multirow{1}{5em}{$Q_{0, -2}Q_{0, 2}$}&$W_{11}^{~~23}W_{23}^{~~11}$\\\hline\multirow{1}{5em}{$Q_{2, -2}Q_{-2, 2}$}&$W_{11}^{~~33}W_{33}^{~~11}$\\\hline\multirow{1}{5em}{$Q_{3, -1}Q_{-3, 1}$}&$W_{12}^{~~33}W_{33}^{~~12}$\\\hline\multirow{1}{5em}{$Q_{4, 0}Q_{-4, 0}$}&$W_{22}^{~~33}W_{33}^{~~22}$\\\hline
        \multirow{27}{16em}{$Q_{1, -1}Q_{1, 1}Q_{-2, 0}$,\hspace{5mm}$Q_{-1, 1}Q_{-1, -1}Q_{2, 0}$}&$(V^I)_{1}^{~3}(V^J)_{2}^{~1}(V^K)_{3}^{~2}$,\hspace{5mm}$(V^I)_{3}^{~1}(V^J)_{1}^{~2}(V^K)_{2}^{~3}$\\&$W_{13}^{~~12}(V^I)_{1}^{~3}(V^J)_{2}^{~1}$\hspace{5mm}$W_{12}^{~~13}(V^I)_{3}^{~1}(V^J)_{1}^{~2}$\\&$W_{23}^{~~22}(V^I)_{1}^{~3}(V^J)_{2}^{~1}$\hspace{5mm}$W_{22}^{~~23}(V^I)_{3}^{~1}(V^J)_{1}^{~2}$\\&$W_{12}^{~~11}(V^I)_{1}^{~3}(V^J)_{3}^{~2}$\hspace{5mm}$W_{11}^{~~12}(V^I)_{3}^{~1}(V^J)_{2}^{~3}$\\&$W_{12}^{~~11}W_{13}^{~~12}(V^I)_{1}^{~3}$\hspace{5mm}$W_{11}^{~~12}W_{12}^{~~13}(V^I)_{3}^{~1}$\\&$W_{12}^{~~11}W_{23}^{~~22}(V^I)_{1}^{~3}$\hspace{5mm}$W_{11}^{~~12}W_{22}^{~~23}(V^I)_{3}^{~1}$\\&$W_{22}^{~~12}(V^I)_{1}^{~3}(V^J)_{3}^{~2}$\hspace{5mm}$W_{12}^{~~22}(V^I)_{3}^{~1}(V^J)_{2}^{~3}$\\&$W_{22}^{~~12}W_{13}^{~~12}(V^I)_{1}^{~3}$\hspace{5mm}$W_{12}^{~~22}W_{12}^{~~13}(V^I)_{3}^{~1}$\\&$W_{22}^{~~12}W_{23}^{~~22}(V^I)_{1}^{~3}$\hspace{5mm}$W_{12}^{~~22}W_{22}^{~~23}(V^I)_{3}^{~1}$\\&$W_{11}^{~~13}(V^I)_{2}^{~1}(V^J)_{3}^{~2}$\hspace{5mm}$W_{13}^{~~11}(V^I)_{1}^{~2}(V^J)_{2}^{~3}$\\&$W_{11}^{~~13}W_{13}^{~~12}(V^I)_{2}^{~1}$\hspace{5mm}$W_{13}^{~~11}W_{12}^{~~13}(V^I)_{1}^{~2}$\\&$W_{11}^{~~13}W_{23}^{~~22}(V^I)_{2}^{~1}$\hspace{5mm}$W_{13}^{~~11}W_{22}^{~~23}(V^I)_{1}^{~2}$\\&$W_{11}^{~~13}W_{12}^{~~11}(V^I)_{3}^{~2}$\hspace{5mm}$W_{13}^{~~11}W_{11}^{~~12}(V^I)_{2}^{~3}$\\&$W_{12}^{~~11}W_{13}^{~~12}W_{11}^{~~13}$\hspace{5mm}$W_{11}^{~~12}W_{12}^{~~13}W_{13}^{~~11}$\\&$W_{12}^{~~11}W_{23}^{~~22}W_{11}^{~~13}$\hspace{5mm}$W_{11}^{~~12}W_{22}^{~~23}W_{13}^{~~11}$\\&$W_{11}^{~~13}W_{22}^{~~12}(V^I)_{3}^{~2}$\hspace{5mm}$W_{13}^{~~11}W_{12}^{~~22}(V^I)_{2}^{~3}$\\&$W_{22}^{~~12}W_{13}^{~~12}W_{11}^{~~13}$\hspace{5mm}$W_{12}^{~~22}W_{12}^{~~13}W_{13}^{~~11}$\\&$W_{22}^{~~12}W_{23}^{~~22}W_{11}^{~~13}$\hspace{5mm}$W_{12}^{~~22}W_{22}^{~~23}W_{13}^{~~11}$\\&$W_{12}^{~~23}(V^I)_{2}^{~1}(V^J)_{3}^{~2}$\hspace{5mm}$W_{23}^{~~12}(V^I)_{1}^{~2}(V^J)_{2}^{~3}$\\&$W_{12}^{~~23}W_{13}^{~~12}(V^I)_{2}^{~1}$\hspace{5mm}$W_{23}^{~~12}W_{12}^{~~13}(V^I)_{1}^{~2}$\\&$W_{12}^{~~23}W_{23}^{~~22}(V^I)_{2}^{~1}$\hspace{5mm}$W_{23}^{~~12}W_{22}^{~~23}(V^I)_{1}^{~2}$\\&$W_{12}^{~~23}W_{12}^{~~11}(V^I)_{3}^{~2}$\hspace{5mm}$W_{23}^{~~12}W_{11}^{~~12}(V^I)_{2}^{~3}$\\&$W_{12}^{~~11}W_{13}^{~~12}W_{12}^{~~23}$\hspace{5mm}$W_{11}^{~~12}W_{12}^{~~13}W_{23}^{~~12}$\\&$W_{12}^{~~11}W_{23}^{~~22}W_{12}^{~~23}$\hspace{5mm}$W_{11}^{~~12}W_{22}^{~~23}W_{23}^{~~12}$\\&$W_{12}^{~~23}W_{22}^{~~12}(V^I)_{3}^{~2}$\hspace{5mm}$W_{23}^{~~12}W_{12}^{~~22}(V^I)_{2}^{~3}$\\&$W_{22}^{~~12}W_{13}^{~~12}W_{12}^{~~23}$\hspace{5mm}$W_{12}^{~~22}W_{12}^{~~13}W_{23}^{~~12}$\\&$W_{22}^{~~12}W_{23}^{~~22}W_{12}^{~~23}$\hspace{5mm}$W_{12}^{~~22}W_{22}^{~~23}W_{23}^{~~12}$\\\hline
\end{tabular}
\end{center}
\begin{center}
\renewcommand{\arraystretch}{1.25}
    \begin{tabular}{|c|c|}
    \hline
        Charges & $U(1)_1\times U(1)_2$ Invariants \\
        \hline
        \multirow{9}{16em}{$Q_{-1, -1}Q_{3, 1}Q_{-2, 0}$,\hspace{5mm}$Q_{1, 1}Q_{-3, -1}Q_{2, 0}$}&$W_{22}^{~~13}(V^I)_{1}^{~2}(V^J)_{3}^{~2}$\hspace{5mm}$W_{13}^{~~22}(V^I)_{2}^{~1}(V^J)_{2}^{~3}$\\&$W_{22}^{~~13}W_{13}^{~~12}(V^I)_{1}^{~2}$\hspace{5mm}$W_{13}^{~~22}W_{12}^{~~13}(V^I)_{2}^{~1}$\\&$W_{22}^{~~13}W_{23}^{~~22}(V^I)_{1}^{~2}$\hspace{5mm}$W_{13}^{~~22}W_{22}^{~~23}(V^I)_{2}^{~1}$\\&$W_{11}^{~~12}W_{22}^{~~13}(V^I)_{3}^{~2}$\hspace{5mm}$W_{12}^{~~11}W_{13}^{~~22}(V^I)_{2}^{~3}$\\&$W_{22}^{~~13}W_{13}^{~~12}W_{11}^{~~12}$\hspace{5mm}$W_{13}^{~~22}W_{12}^{~~13}W_{12}^{~~11}$\\&$W_{22}^{~~13}W_{23}^{~~22}W_{11}^{~~12}$\hspace{5mm}$W_{13}^{~~22}W_{22}^{~~23}W_{12}^{~~11}$\\&$W_{12}^{~~22}W_{22}^{~~13}(V^I)_{3}^{~2}$\hspace{5mm}$W_{22}^{~~12}W_{13}^{~~22}(V^I)_{2}^{~3}$\\&$W_{22}^{~~13}W_{13}^{~~12}W_{12}^{~~22}$\hspace{5mm}$W_{13}^{~~22}W_{12}^{~~13}W_{22}^{~~12}$\\&$W_{22}^{~~13}W_{23}^{~~22}W_{12}^{~~22}$\hspace{5mm}$W_{13}^{~~22}W_{22}^{~~23}W_{22}^{~~12}$\\\hline\multirow{9}{16em}{$Q_{0, -2}Q_{1, 1}Q_{-1, 1}$,\hspace{5mm}$Q_{0, 2}Q_{-1, -1}Q_{1, -1}$}&$W_{11}^{~~23}(V^I)_{2}^{~1}(V^J)_{3}^{~1}$\hspace{5mm}$W_{23}^{~~11}(V^I)_{1}^{~2}(V^J)_{1}^{~3}$\\&$W_{11}^{~~23}W_{13}^{~~11}(V^I)_{2}^{~1}$\hspace{5mm}$W_{23}^{~~11}W_{11}^{~~13}(V^I)_{1}^{~2}$\\&$W_{11}^{~~23}W_{23}^{~~12}(V^I)_{2}^{~1}$\hspace{5mm}$W_{23}^{~~11}W_{12}^{~~23}(V^I)_{1}^{~2}$\\&$W_{11}^{~~23}W_{12}^{~~11}(V^I)_{3}^{~1}$\hspace{5mm}$W_{23}^{~~11}W_{11}^{~~12}(V^I)_{1}^{~3}$\\&$W_{12}^{~~11}W_{13}^{~~11}W_{11}^{~~23}$\hspace{5mm}$W_{11}^{~~12}W_{11}^{~~13}W_{23}^{~~11}$\\&$W_{12}^{~~11}W_{23}^{~~12}W_{11}^{~~23}$\hspace{5mm}$W_{11}^{~~12}W_{12}^{~~23}W_{23}^{~~11}$\\&$W_{11}^{~~23}W_{22}^{~~12}(V^I)_{3}^{~1}$\hspace{5mm}$W_{23}^{~~11}W_{12}^{~~22}(V^I)_{1}^{~3}$\\&$W_{22}^{~~12}W_{13}^{~~11}W_{11}^{~~23}$\hspace{5mm}$W_{12}^{~~22}W_{11}^{~~13}W_{23}^{~~11}$\\&$W_{22}^{~~12}W_{23}^{~~12}W_{11}^{~~23}$\hspace{5mm}$W_{12}^{~~22}W_{12}^{~~23}W_{23}^{~~11}$\\\hline\multirow{9}{16em}{$Q_{2, 0}Q_{1, -1}Q_{-3, 1}$,\hspace{5mm}$Q_{-2, 0}Q_{-1, 1}Q_{3, -1}$}&$W_{33}^{~~12}(V^I)_{2}^{~3}(V^J)_{1}^{~3}$\hspace{5mm}$W_{12}^{~~33}(V^I)_{3}^{~2}(V^J)_{3}^{~1}$\\&$W_{11}^{~~13}W_{33}^{~~12}(V^I)_{2}^{~3}$\hspace{5mm}$W_{13}^{~~11}W_{12}^{~~33}(V^I)_{3}^{~2}$\\&$W_{12}^{~~23}W_{33}^{~~12}(V^I)_{2}^{~3}$\hspace{5mm}$W_{23}^{~~12}W_{12}^{~~33}(V^I)_{3}^{~2}$\\&$W_{12}^{~~13}W_{33}^{~~12}(V^I)_{1}^{~3}$\hspace{5mm}$W_{13}^{~~12}W_{12}^{~~33}(V^I)_{3}^{~1}$\\&$W_{11}^{~~13}W_{33}^{~~12}W_{12}^{~~13}$\hspace{5mm}$W_{13}^{~~11}W_{12}^{~~33}W_{13}^{~~12}$\\&$W_{12}^{~~23}W_{33}^{~~12}W_{12}^{~~13}$\hspace{5mm}$W_{23}^{~~12}W_{12}^{~~33}W_{13}^{~~12}$\\&$W_{22}^{~~23}W_{33}^{~~12}(V^I)_{1}^{~3}$\hspace{5mm}$W_{23}^{~~22}W_{12}^{~~33}(V^I)_{3}^{~1}$\\&$W_{11}^{~~13}W_{33}^{~~12}W_{22}^{~~23}$\hspace{5mm}$W_{13}^{~~11}W_{12}^{~~33}W_{23}^{~~22}$\\&$W_{12}^{~~23}W_{33}^{~~12}W_{22}^{~~23}$\hspace{5mm}$W_{23}^{~~12}W_{12}^{~~33}W_{23}^{~~22}$\\\hline
        \multirow{6}{16em}{$Q_{-4, 0}Q_{2, 0}Q_{2, 0}$\hspace{5mm}$Q_{4, 0}Q_{-2, 0}Q_{-2, 0}$}&$W_{33}^{~~22}(V^I)_{2}^{~3}(V^J)_{2}^{~3}$\hspace{5mm}$W_{22}^{~~33}(V^I)_{3}^{~2}(V^J)_{3}^{~2}$\\&$W_{33}^{~~22}W_{12}^{~~13}(V^I)_{2}^{~3}$\hspace{5mm}$W_{22}^{~~33}W_{13}^{~~12}(V^I)_{3}^{~2}$\\&$W_{33}^{~~22}W_{22}^{~~23}(V^I)_{2}^{~3}$\hspace{5mm}$W_{22}^{~~33}W_{23}^{~~22}(V^I)_{3}^{~2}$\\&$W_{12}^{~~13}W_{12}^{~~13}W_{33}^{~~22}$\hspace{5mm}$W_{13}^{~~12}W_{13}^{~~12}W_{22}^{~~33}$\\&$W_{12}^{~~13}W_{22}^{~~23}W_{33}^{~~22}$\hspace{5mm}$W_{13}^{~~12}W_{23}^{~~22}W_{22}^{~~33}$\\&$W_{22}^{~~23}W_{22}^{~~23}W_{33}^{~~22}$\hspace{5mm}$W_{23}^{~~22}W_{23}^{~~22}W_{22}^{~~33}$\\\hline
\end{tabular}
\end{center}
\begin{center}
\renewcommand{\arraystretch}{1.25}
    \begin{tabular}{|c|c|}
    \hline
        Charges & $U(1)_1\times U(1)_2$ Invariants \\
        \hline
        \multirow{6}{16em}{$Q_{-2, -2}Q_{1, 1}Q_{1, 1}$\hspace{5mm}$Q_{2, 2}Q_{-1, -1}Q_{-1, -1}$}&$W_{11}^{~~22}(V^I)_{2}^{~1}(V^J)_{2}^{~1}$\hspace{5mm}$W_{22}^{~~11}(V^I)_{1}^{~2}(V^J)_{1}^{~2}$\\&$W_{11}^{~~22}W_{12}^{~~11}(V^I)_{2}^{~1}$\hspace{5mm}$W_{22}^{~~11}W_{11}^{~~12}(V^I)_{1}^{~2}$\\&$W_{11}^{~~22}W_{22}^{~~12}(V^I)_{2}^{~1}$\hspace{5mm}$W_{22}^{~~11}W_{12}^{~~22}(V^I)_{1}^{~2}$\\&$W_{12}^{~~11}W_{12}^{~~11}W_{11}^{~~22}$\hspace{5mm}$W_{11}^{~~12}W_{11}^{~~12}W_{22}^{~~11}$\\&$W_{12}^{~~11}W_{22}^{~~12}W_{11}^{~~22}$\hspace{5mm}$W_{11}^{~~12}W_{12}^{~~22}W_{22}^{~~11}$\\&$W_{22}^{~~12}W_{22}^{~~12}W_{11}^{~~22}$\hspace{5mm}$W_{12}^{~~22}W_{12}^{~~22}W_{22}^{~~11}$\\\hline\multirow{6}{16em}{$Q_{-2, 2}Q_{1, -1}Q_{1, -1}$\hspace{5mm}$Q_{2, -2}Q_{-1, 1}Q_{-1, 1}$}&$W_{33}^{~~11}(V^I)_{1}^{~3}(V^J)_{1}^{~3}$\hspace{5mm}$W_{11}^{~~33}(V^I)_{3}^{~1}(V^J)_{3}^{~1}$\\&$W_{33}^{~~11}W_{11}^{~~13}(V^I)_{1}^{~3}$\hspace{5mm}$W_{11}^{~~33}W_{13}^{~~11}(V^I)_{3}^{~1}$\\&$W_{33}^{~~11}W_{12}^{~~23}(V^I)_{1}^{~3}$\hspace{5mm}$W_{11}^{~~33}W_{23}^{~~12}(V^I)_{3}^{~1}$\\&$W_{11}^{~~13}W_{11}^{~~13}W_{33}^{~~11}$\hspace{5mm}$W_{13}^{~~11}W_{13}^{~~11}W_{11}^{~~33}$\\&$W_{11}^{~~13}W_{12}^{~~23}W_{33}^{~~11}$\hspace{5mm}$W_{13}^{~~11}W_{23}^{~~12}W_{11}^{~~33}$\\&$W_{12}^{~~23}W_{12}^{~~23}W_{33}^{~~11}$\hspace{5mm}$W_{23}^{~~12}W_{23}^{~~12}W_{11}^{~~33}$\\\hline
        \multirow{3}{16em}{$Q_{-2, -2}Q_{3, 1}Q_{-1, 1}$,\hspace{5mm}$Q_{2, 2}Q_{-3, -1}Q_{1, -1}$}&$W_{11}^{~~22}W_{22}^{~~13}(V^I)_{3}^{~1}$\hspace{5mm}$W_{22}^{~~11}W_{13}^{~~22}(V^I)_{1}^{~3}$\\&$W_{22}^{~~13}W_{13}^{~~11}W_{11}^{~~22}$\hspace{5mm}$W_{13}^{~~22}W_{11}^{~~13}W_{22}^{~~11}$\\&$W_{22}^{~~13}W_{23}^{~~12}W_{11}^{~~22}$\hspace{5mm}$W_{13}^{~~22}W_{12}^{~~23}W_{22}^{~~11}$\\\hline
        \multirow{3}{16em}{$Q_{-2, 2}Q_{3, -1}Q_{-1, -1}$,\hspace{5mm}$Q_{2, -2}Q_{-3, 1}Q_{1, 1}$}&$W_{33}^{~~11}W_{12}^{~~33}(V^I)_{1}^{~2}$\hspace{5mm}$W_{11}^{~~33}W_{33}^{~~12}(V^I)_{2}^{~1}$\\&$W_{12}^{~~33}W_{11}^{~~12}W_{33}^{~~11}$\hspace{5mm}$W_{33}^{~~12}W_{12}^{~~11}W_{11}^{~~33}$\\&$W_{12}^{~~33}W_{12}^{~~22}W_{33}^{~~11}$\hspace{5mm}$W_{33}^{~~12}W_{22}^{~~12}W_{11}^{~~33}$\\\hline\multirow{3}{16em}{$Q_{0, -2}Q_{2, 2}Q_{-2, 0}$,\hspace{5mm}$Q_{0, 2}Q_{-2, -2}Q_{2, 0}$}&$W_{11}^{~~23}W_{22}^{~~11}(V^I)_{3}^{~2}$\hspace{5mm}$W_{23}^{~~11}W_{11}^{~~22}(V^I)_{2}^{~3}$\\&$W_{22}^{~~11}W_{13}^{~~12}W_{11}^{~~23}$\hspace{5mm}$W_{11}^{~~22}W_{12}^{~~13}W_{23}^{~~11}$\\&$W_{22}^{~~11}W_{23}^{~~22}W_{11}^{~~23}$\hspace{5mm}$W_{11}^{~~22}W_{22}^{~~23}W_{23}^{~~11}$\\\hline\multirow{3}{16em}{$Q_{1, -1}Q_{3, 1}Q_{-4, 0}$,\hspace{5mm}$Q_{-1, 1}Q_{-3, -1}Q_{4, 0}$}&$W_{22}^{~~13}W_{33}^{~~22}(V^I)_{1}^{~3}$\hspace{5mm}$W_{13}^{~~22}W_{22}^{~~33}(V^I)_{3}^{~1}$\\&$W_{22}^{~~13}W_{33}^{~~22}W_{11}^{~~13}$\hspace{5mm}$W_{13}^{~~22}W_{22}^{~~33}W_{13}^{~~11}$\\&$W_{22}^{~~13}W_{33}^{~~22}W_{12}^{~~23}$\hspace{5mm}$W_{13}^{~~22}W_{22}^{~~33}W_{23}^{~~12}$\\\hline\multirow{3}{16em}{$Q_{2, -2}Q_{0, 2}Q_{-2, 0}$,\hspace{5mm}$Q_{-2, 2}Q_{0, -2}Q_{2, 0}$}&$W_{11}^{~~33}W_{23}^{~~11}(V^I)_{3}^{~2}$\hspace{5mm}$W_{33}^{~~11}W_{11}^{~~23}(V^I)_{2}^{~3}$\\&$W_{23}^{~~11}W_{13}^{~~12}W_{11}^{~~33}$\hspace{5mm}$W_{11}^{~~23}W_{12}^{~~13}W_{33}^{~~11}$\\&$W_{23}^{~~11}W_{23}^{~~22}W_{11}^{~~33}$\hspace{5mm}$W_{11}^{~~23}W_{22}^{~~23}W_{33}^{~~11}$\\\hline\multirow{3}{16em}{$Q_{3, -1}Q_{1, 1}Q_{-4, 0}$,\hspace{5mm}$Q_{-3, 1}Q_{-1, -1}Q_{4, 0}$}&$W_{12}^{~~33}W_{33}^{~~22}(V^I)_{2}^{~1}$\hspace{5mm}$W_{33}^{~~12}W_{22}^{~~33}(V^I)_{1}^{~2}$\\&$W_{12}^{~~11}W_{33}^{~~22}W_{12}^{~~33}$\hspace{5mm}$W_{11}^{~~12}W_{22}^{~~33}W_{33}^{~~12}$\\&$W_{22}^{~~12}W_{33}^{~~22}W_{12}^{~~33}$\hspace{5mm}$W_{12}^{~~22}W_{22}^{~~33}W_{33}^{~~12}$\\\hline\multirow{1}{16em}{$Q_{0, -2}Q_{3, 1}Q_{-3, 1}$,\hspace{5mm}$Q_{0, 2}Q_{-3, -1}Q_{3, -1}$}&$W_{22}^{~~13}W_{33}^{~~12}W_{11}^{~~23}$\hspace{5mm}$W_{13}^{~~22}W_{12}^{~~33}W_{23}^{~~11}$\\\hline\multirow{1}{16em}{$Q_{2, -2}Q_{2, 2}Q_{-4, 0}$,\hspace{5mm}$Q_{-2, 2}Q_{-2, -2}Q_{4, 0}$}&$W_{22}^{~~11}W_{33}^{~~22}W_{11}^{~~33}$\hspace{5mm}$W_{11}^{~~22}W_{22}^{~~33}W_{33}^{~~11}$\\\hline
\end{tabular}
\end{center}

We next list the remaining $U(1)_1\times U(1)_2$ invariants to arbitrary order in $\lambda$. To conserve space, these invariants are presented only in terms of their field charges rather than the corresponding field components. Furthermore, we list only one representative from each pair of invariants related by the transformation
$Q_{q_1,q_2}\rightarrow Q_{-q_1,-q_2}$.
The omitted invariants are obtained by applying this transformation to every factor appearing in the listed invariant.
\begin{center}
\parbox{\textwidth}{
\centering
{$\mathcal{O}(\lambda^4)$ $U(1)_1\times U(1)_2$ invariants}

\vspace{2 mm}

\scriptsize
\renewcommand{\arraystretch}{1.3}
\begin{tabular}{ccc}
\hline
$(Q_{0,-2})^{2}Q_{2,2}Q_{-2,2}$ &$Q_{0,-2}Q_{2,2}Q_{-3,-1}Q_{1,1}$ &  $Q_{0,-2}Q_{2,2}Q_{1,-1}Q_{-3,1}$  \\ [1mm]$Q_{0,-2}(Q_{1,1})^{2}Q_{-2,0}$ &$Q_{0,-2}Q_{3,1}Q_{-1,1}Q_{-2,0}$ &  $Q_{2,-2}Q_{0,2}Q_{-3,-1}Q_{1,1}$  \\ [1mm]$Q_{2,-2}Q_{0,2}Q_{1,-1}Q_{-3,1}$ &$Q_{2,-2}Q_{3,1}Q_{-3,1}Q_{-2,0}$ &  $(Q_{-1,-1})^{2}Q_{3,1}Q_{-1,1}$  \\ [1mm]$(Q_{1,-1})^{2}Q_{1,1}Q_{-3,1}$ &$Q_{2,0}Q_{-2,-2}Q_{3,1}Q_{-3,1}$ &  $Q_{2,0}Q_{0,-2}(Q_{-1,1})^{2}$  \\ [1mm]$Q_{2,0}Q_{0,-2}Q_{1,1}Q_{-3,1}$ &$Q_{-2,-2}Q_{-2,2}(Q_{2,0})^{2}$ &  $Q_{-3,-1}Q_{-1,1}(Q_{2,0})^{2}$  \\ [1mm]$Q_{-1,-1}Q_{-3,1}(Q_{2,0})^{2}$ &$Q_{-4,0}Q_{2,0}Q_{0,-2}Q_{2,2}$ &  $Q_{-4,0}Q_{2,0}Q_{2,-2}Q_{0,2}$  \\ [1mm]$Q_{-4,0}Q_{2,0}Q_{-1,-1}Q_{3,1}$ &$Q_{-4,0}Q_{2,0}Q_{1,-1}Q_{1,1}$ &  $Q_{-4,0}Q_{2,0}Q_{3,-1}Q_{-1,1}$  \\ [1mm]$Q_{-2,-2}(Q_{3,1})^{2}Q_{-4,0}$ &$Q_{0,-2}Q_{3,1}Q_{1,1}Q_{-4,0}$ &  $Q_{2,-2}(Q_{1,1})^{2}Q_{-4,0}$  \\ [1mm]$Q_{1,-1}Q_{3,1}Q_{-1,-1}Q_{-3,1}$ &$Q_{4,0}Q_{-2,-2}(Q_{-1,1})^{2}$ &  $Q_{4,0}Q_{0,-2}Q_{-1,1}Q_{-3,1}$  \\ [1mm]$Q_{4,0}Q_{2,-2}(Q_{-3,1})^{2}$ &  $Q_{3,-1}Q_{3,1}Q_{-4,0}Q_{-2,0}$ & \\ [1mm]
\hline
\end{tabular}
}
\end{center}
\begin{center}
\parbox{\textwidth}{
\centering
{$\mathcal{O}(\lambda^5)$ $U(1)_1\times U(1)_2$ invariants}

\vspace{2 mm}

\scriptsize
\renewcommand{\arraystretch}{1.3}
\begin{tabular}{ccc}
\hline
$(Q_{0,-2})^{2}Q_{2,2}(Q_{-1,1})^{2}$ &$(Q_{0,-2})^{2}Q_{2,2}Q_{1,1}Q_{-3,1}$ &  $(Q_{0,-2})^{2}(Q_{1,1})^{2}Q_{-2,2}$  \\ [1mm]$Q_{0,-2}(Q_{1,1})^{3}Q_{-3,-1}$ &$(Q_{0,-2})^{2}Q_{3,1}Q_{-1,1}Q_{-2,2}$ &  $Q_{2,-2}Q_{3,1}Q_{-3,1}Q_{-2,-2}Q_{0,2}$  \\ [1mm]$Q_{-1,-1}Q_{3,1}Q_{0,-2}(Q_{-1,1})^{2}$ &$Q_{1,-1}(Q_{1,1})^{2}Q_{0,-2}Q_{-3,1}$ &  $Q_{3,-1}(Q_{-1,1})^{3}Q_{0,-2}$  \\ [1mm]$Q_{-2,-2}Q_{-2,2}Q_{2,0}Q_{-1,-1}Q_{3,1}$ &$Q_{-2,-2}Q_{-2,2}Q_{2,0}Q_{3,-1}Q_{-1,1}$ &  $Q_{-2,-2}(Q_{-1,1})^{2}(Q_{2,0})^{2}$  \\ [1mm]$Q_{0,-2}Q_{-1,1}Q_{-3,1}(Q_{2,0})^{2}$ &$Q_{2,-2}(Q_{-3,1})^{2}(Q_{2,0})^{2}$ &  $Q_{-3,-1}Q_{-1,1}Q_{2,0}Q_{0,-2}Q_{2,2}$  \\ [1mm]$Q_{-3,-1}(Q_{-1,1})^{2}Q_{2,0}Q_{3,-1}$ &$Q_{-1,-1}Q_{-3,1}Q_{2,0}Q_{2,-2}Q_{0,2}$ &  $(Q_{-1,-1})^{2}Q_{-3,1}Q_{2,0}Q_{3,1}$  \\ [1mm]$Q_{-4,0}(Q_{0,-2})^{2}(Q_{2,2})^{2}$ &$Q_{-4,0}(Q_{2,-2})^{2}(Q_{0,2})^{2}$ &  $Q_{-4,0}Q_{-1,-1}Q_{3,1}Q_{2,-2}Q_{0,2}$  \\ [1mm]$Q_{-4,0}(Q_{-1,-1})^{2}(Q_{3,1})^{2}$ &$Q_{-4,0}Q_{1,-1}Q_{1,1}Q_{0,-2}Q_{2,2}$ &  $Q_{-4,0}Q_{1,-1}Q_{1,1}Q_{2,-2}Q_{0,2}$  \\ [1mm]$Q_{-4,0}(Q_{1,-1})^{2}(Q_{1,1})^{2}$ &$Q_{-4,0}Q_{3,-1}Q_{-1,1}Q_{0,-2}Q_{2,2}$ &  $Q_{-4,0}(Q_{3,-1})^{2}(Q_{-1,1})^{2}$  \\ [1mm]$Q_{-4,0}Q_{2,0}Q_{0,-2}(Q_{1,1})^{2}$ &$Q_{-4,0}Q_{2,0}Q_{0,-2}Q_{3,1}Q_{-1,1}$ &  $Q_{-4,0}Q_{2,0}Q_{2,-2}Q_{3,1}Q_{-3,1}$  \\ [1mm]$Q_{-4,0}Q_{0,2}(Q_{1,-1})^{2}Q_{2,0}$ &$Q_{-4,0}Q_{0,2}Q_{3,-1}Q_{-1,-1}Q_{2,0}$ &  $Q_{-4,0}Q_{2,2}Q_{3,-1}Q_{-3,-1}Q_{2,0}$  \\ [1mm]$Q_{-2,2}(Q_{-1,-1})^{2}(Q_{2,0})^{2}$ &$Q_{0,2}Q_{-1,-1}Q_{-3,-1}(Q_{2,0})^{2}$ &  $Q_{2,2}(Q_{-3,-1})^{2}(Q_{2,0})^{2}$  \\ [1mm]$(Q_{-2,-2})^{2}(Q_{3,1})^{2}Q_{-2,2}$ &$Q_{-2,-2}(Q_{3,1})^{2}Q_{-1,-1}Q_{-3,1}$ &  $(Q_{2,-2})^{2}Q_{2,2}(Q_{-3,1})^{2}$  \\ [1mm]$Q_{1,-1}Q_{3,1}Q_{2,-2}(Q_{-3,1})^{2}$ &$Q_{-3,-1}Q_{-3,1}(Q_{2,0})^{3}$ &  $Q_{0,-2}(Q_{3,1})^{2}Q_{-4,0}Q_{-2,0}$  \\ [1mm]$Q_{2,-2}Q_{3,1}Q_{1,1}Q_{-4,0}Q_{-2,0}$ &$Q_{3,-1}Q_{3,1}Q_{-2,-2}Q_{-2,2}Q_{-2,0}$ &  $Q_{3,-1}Q_{3,1}Q_{-4,0}Q_{-2,-2}Q_{0,2}$  \\ [1mm]$Q_{3,-1}Q_{3,1}Q_{-4,0}Q_{0,-2}Q_{-2,2}$ &$Q_{3,-1}Q_{3,1}Q_{-4,0}Q_{-1,-1}Q_{-1,1}$ &  $Q_{0,2}(Q_{3,-1})^{2}Q_{-4,0}Q_{-2,0}$  \\ [1mm]$Q_{2,2}Q_{3,-1}Q_{1,-1}Q_{-4,0}Q_{-2,0}$ &$Q_{-2,-2}(Q_{-3,1})^{2}(Q_{4,0})^{2}$ &  $Q_{-2,2}(Q_{-3,-1})^{2}(Q_{4,0})^{2}$  \\ [1mm]
\hline
\end{tabular}
}
\end{center}
\begin{center}
\parbox{\textwidth}{
\centering
{$\mathcal{O}(\lambda^6)$ $U(1)_1\times U(1)_2$ invariants}

\vspace{2 mm}

\scriptsize
\renewcommand{\arraystretch}{1.3}
\begin{tabular}{ccc}
\hline
$(Q_{0,-2})^{2}(Q_{1,1})^{3}Q_{-3,1}$ &$(Q_{0,-2})^{2}Q_{3,1}(Q_{-1,1})^{3}$ &  $Q_{-2,-2}(Q_{-1,1})^{3}Q_{2,0}Q_{3,-1}$  \\ [1mm]$(Q_{0,-2})^{2}Q_{-1,1}Q_{-3,1}Q_{2,0}Q_{2,2}$ &$(Q_{2,-2})^{2}(Q_{-3,1})^{2}Q_{2,0}Q_{0,2}$ &  $Q_{-3,-1}Q_{-1,1}(Q_{0,-2})^{2}(Q_{2,2})^{2}$  \\ [1mm]$Q_{-3,-1}(Q_{-1,1})^{2}Q_{3,-1}Q_{0,-2}Q_{2,2}$ &$Q_{-3,-1}(Q_{-1,1})^{2}Q_{3,-1}Q_{0,-2}Q_{2,2}$ &  $Q_{-3,-1}(Q_{-1,1})^{3}(Q_{3,-1})^{2}$  \\ [1mm]$Q_{-1,-1}Q_{-3,1}(Q_{2,-2})^{2}(Q_{0,2})^{2}$ &$(Q_{-1,-1})^{2}Q_{-3,1}Q_{3,1}Q_{2,-2}Q_{0,2}$ &  $(Q_{-1,-1})^{2}Q_{-3,1}Q_{3,1}Q_{2,-2}Q_{0,2}$  \\ [1mm]$(Q_{-1,-1})^{3}Q_{-3,1}(Q_{3,1})^{2}$ &$Q_{-1,-1}(Q_{-3,1})^{2}Q_{2,0}Q_{2,-2}Q_{3,1}$ &  $Q_{-4,0}(Q_{0,-2})^{2}(Q_{1,1})^{2}Q_{2,2}$  \\ [1mm]$Q_{-4,0}(Q_{0,-2})^{2}Q_{3,1}Q_{-1,1}Q_{2,2}$ &$Q_{-4,0}(Q_{2,-2})^{2}Q_{3,1}Q_{-3,1}Q_{0,2}$ &  $Q_{-4,0}Q_{1,-1}(Q_{1,1})^{3}Q_{0,-2}$  \\ [1mm]$Q_{-4,0}(Q_{0,2})^{2}(Q_{1,-1})^{2}Q_{2,-2}$ &$Q_{-4,0}Q_{0,2}(Q_{1,-1})^{3}Q_{1,1}$ &  $Q_{-4,0}(Q_{0,2})^{2}Q_{3,-1}Q_{-1,-1}Q_{2,-2}$  \\ [1mm]$Q_{-4,0}(Q_{2,2})^{2}Q_{3,-1}Q_{-3,-1}Q_{0,-2}$ &$Q_{-2,2}(Q_{-1,-1})^{3}Q_{2,0}Q_{3,1}$ &  $(Q_{0,2})^{2}Q_{-1,-1}Q_{-3,-1}Q_{2,0}Q_{2,-2}$  \\ [1mm]$(Q_{2,2})^{2}(Q_{-3,-1})^{2}Q_{2,0}Q_{0,-2}$ &$Q_{2,2}(Q_{-3,-1})^{2}Q_{2,0}Q_{3,-1}Q_{-1,1}$ &  $Q_{-2,-2}(Q_{3,1})^{2}Q_{-2,2}(Q_{-1,-1})^{2}$  \\ [1mm]$Q_{-2,2}(Q_{3,-1})^{2}Q_{-2,-2}(Q_{-1,1})^{2}$ &$Q_{-2,-2}Q_{-1,1}Q_{-3,1}(Q_{2,0})^{3}$ &  $Q_{0,-2}(Q_{-3,1})^{2}(Q_{2,0})^{3}$  \\ [1mm]$Q_{-3,-1}Q_{-3,1}(Q_{2,0})^{2}Q_{0,-2}Q_{2,2}$ &$Q_{-3,-1}Q_{-3,1}(Q_{2,0})^{2}Q_{0,-2}Q_{2,2}$ &  $Q_{-3,-1}Q_{-3,1}(Q_{2,0})^{2}Q_{2,-2}Q_{0,2}$  \\ [1mm]$Q_{-3,-1}Q_{-3,1}(Q_{2,0})^{2}Q_{2,-2}Q_{0,2}$ &$Q_{-2,2}Q_{-1,-1}Q_{-3,-1}(Q_{2,0})^{3}$ &  $Q_{0,2}(Q_{-3,-1})^{2}(Q_{2,0})^{3}$  \\ [1mm]$Q_{0,-2}(Q_{3,1})^{2}Q_{-2,-2}Q_{-2,2}Q_{-2,0}$ &$(Q_{0,-2})^{2}(Q_{3,1})^{2}Q_{-4,0}Q_{-2,2}$ &  $Q_{0,-2}(Q_{3,1})^{2}Q_{-4,0}Q_{-1,-1}Q_{-1,1}$  \\ [1mm]$Q_{3,-1}Q_{3,1}(Q_{-2,-2})^{2}Q_{-2,2}Q_{0,2}$ &$Q_{3,-1}Q_{3,1}Q_{-2,-2}(Q_{-2,2})^{2}Q_{0,-2}$ &  $Q_{3,-1}Q_{3,1}Q_{-4,0}Q_{0,-2}(Q_{-1,1})^{2}$  \\ [1mm]$Q_{3,-1}Q_{3,1}Q_{-4,0}Q_{0,2}(Q_{-1,-1})^{2}$ &$Q_{0,2}(Q_{3,-1})^{2}Q_{-2,-2}Q_{-2,2}Q_{-2,0}$ &  $(Q_{0,2})^{2}(Q_{3,-1})^{2}Q_{-4,0}Q_{-2,-2}$  \\ [1mm]$Q_{0,2}(Q_{3,-1})^{2}Q_{-4,0}Q_{-1,-1}Q_{-1,1}$ &$Q_{-2,-2}(Q_{-3,1})^{2}Q_{4,0}(Q_{2,0})^{2}$ &  $Q_{-2,2}(Q_{-3,-1})^{2}Q_{4,0}(Q_{2,0})^{2}$  \\ [1mm]$Q_{-2,-2}(Q_{-3,1})^{2}Q_{4,0}Q_{1,-1}Q_{3,1}$ &$Q_{-2,2}(Q_{-3,-1})^{2}Q_{4,0}Q_{3,-1}Q_{1,1}$ &  $Q_{2,-2}(Q_{3,1})^{2}Q_{-2,-2}(Q_{-3,1})^{2}$  \\ [1mm]
\hline
\end{tabular}
}
\end{center}
\begin{center}
\parbox{\textwidth}{
\centering
{$\mathcal{O}(\lambda^7)$ $U(1)_1\times U(1)_2$ invariants}

\vspace{2 mm}

\scriptsize
\renewcommand{\arraystretch}{1.3}
\begin{tabular}{ccc}
\hline
$Q_{-2,-2}(Q_{-1,1})^{4}(Q_{3,-1})^{2}$ &$(Q_{0,-2})^{3}Q_{-1,1}Q_{-3,1}(Q_{2,2})^{2}$ &  $(Q_{2,-2})^{3}(Q_{-3,1})^{2}(Q_{0,2})^{2}$  \\ [1mm]$(Q_{2,-2})^{2}(Q_{-3,1})^{3}Q_{2,0}Q_{3,1}$ &$Q_{-4,0}(Q_{0,-2})^{2}(Q_{1,1})^{4}$ &  $Q_{-4,0}(Q_{0,2})^{2}(Q_{1,-1})^{4}$  \\ [1mm]$Q_{-2,2}(Q_{-1,-1})^{4}(Q_{3,1})^{2}$ &$(Q_{0,2})^{3}Q_{-1,-1}Q_{-3,-1}(Q_{2,-2})^{2}$ &  $(Q_{2,2})^{3}(Q_{-3,-1})^{2}(Q_{0,-2})^{2}$  \\ [1mm]$(Q_{2,2})^{2}(Q_{-3,-1})^{3}Q_{3,-1}Q_{2,0}$ &$(Q_{0,-2})^{2}(Q_{-3,1})^{2}(Q_{2,0})^{2}Q_{2,2}$ &  $(Q_{0,-2})^{2}(Q_{-3,1})^{2}(Q_{2,0})^{2}Q_{2,2}$  \\ [1mm]$(Q_{0,2})^{2}(Q_{-3,-1})^{2}(Q_{2,0})^{2}Q_{2,-2}$ &$(Q_{0,2})^{2}(Q_{-3,-1})^{2}(Q_{2,0})^{2}Q_{2,-2}$ &  $(Q_{0,-2})^{2}(Q_{3,1})^{2}Q_{-2,-2}(Q_{-2,2})^{2}$  \\ [1mm]$(Q_{0,-2})^{2}(Q_{3,1})^{2}Q_{-4,0}(Q_{-1,1})^{2}$ &$(Q_{0,2})^{2}(Q_{3,-1})^{2}(Q_{-2,-2})^{2}Q_{-2,2}$ &  $(Q_{0,2})^{2}(Q_{3,-1})^{2}Q_{-4,0}(Q_{-1,-1})^{2}$  \\ [1mm]$Q_{-2,-2}(Q_{-3,1})^{2}(Q_{2,0})^{4}$ &$Q_{-2,2}(Q_{-3,-1})^{2}(Q_{2,0})^{4}$ &  $Q_{-2,-2}(Q_{-3,1})^{2}(Q_{1,-1})^{2}(Q_{3,1})^{2}$  \\ [1mm]$Q_{-2,-2}(Q_{-3,1})^{2}(Q_{1,-1})^{2}(Q_{3,1})^{2}$ &$(Q_{-2,-2})^{2}(Q_{-3,1})^{2}Q_{4,0}(Q_{3,1})^{2}$ &  $Q_{-2,2}(Q_{-3,-1})^{2}(Q_{3,-1})^{2}(Q_{1,1})^{2}$  \\ [1mm]$Q_{-2,2}(Q_{-3,-1})^{2}(Q_{3,-1})^{2}(Q_{1,1})^{2}$ &  $(Q_{-2,2})^{2}(Q_{-3,-1})^{2}(Q_{3,-1})^{2}Q_{4,0}$  \\ [1mm]
\hline
\end{tabular}
}
\end{center}
\begin{center}
\parbox{\textwidth}{
\centering
{$\mathcal{O}(\lambda^8)$ $U(1)_1\times U(1)_2$ invariants}

\vspace{2 mm}

\scriptsize
\renewcommand{\arraystretch}{1.3}
\begin{tabular}{ccc}
\hline
$(Q_{2,-2})^{3}(Q_{-3,1})^{3}Q_{3,1}Q_{0,2}$ &$(Q_{2,2})^{3}(Q_{-3,-1})^{3}Q_{3,-1}Q_{0,-2}$ &  $Q_{-3,-1}Q_{-3,1}(Q_{0,-2})^{3}(Q_{2,2})^{3}$  \\ [1mm]  $Q_{-3,-1}Q_{-3,1}(Q_{2,-2})^{3}(Q_{0,2})^{3}$  \\ [1mm]
\hline
\end{tabular}
}
\end{center}
\begin{center}
\parbox{\textwidth}{
\centering
{$\mathcal{O}(\lambda^9)$ $U(1)_1\times U(1)_2$ invariants}

\vspace{2 mm}

\scriptsize
\renewcommand{\arraystretch}{1.3}
\begin{tabular}{ccc}
\hline
$(Q_{0,-2})^{4}(Q_{-3,1})^{2}(Q_{2,2})^{3}$ &$(Q_{0,2})^{4}(Q_{-3,-1})^{2}(Q_{2,-2})^{3}$ &  $(Q_{-2,-2})^{3}(Q_{-3,1})^{2}(Q_{3,1})^{4}$  \\ [1mm]  $(Q_{-2,2})^{3}(Q_{-3,-1})^{2}(Q_{3,-1})^{4}$  \\ [1mm]
\hline
\end{tabular}
}
\end{center}

\clearpage

\bibliographystyle{JHEP}

\end{document}